\documentclass[prb,aps,twocolumn,floats,showpacs,amsmath]{revtex4}

\usepackage[pdftex]{color,graphicx}

\usepackage{amssymb}

\usepackage{color}
\definecolor{blue}{rgb}{0,0,1}
\definecolor{grey}{rgb}{0.6,0.6,0.6}

\def	\bse{\begin{subequations}}
\def	\ese{\end{subequations}}

\usepackage{multirow}

\newcommand{\be}{\begin{equation}}
\newcommand{\ee}{\end{equation}}
\newcommand{\bea}{\begin{eqnarray}}
\newcommand{\eea}{\end{eqnarray}}

\def \kB{k_{\rm B}}

\def \bpsi{\mathbf{\Psi}}
\def \br{\mathbf{r}}
\def \bp{\mathbf{p}}
\def \bk{\mathbf{k}}
\def \bq{\mathbf{q}}

\def \hSigma{\hat{\Sigma}}

\def \HH{\mathcal{H}}

\def \ve{\varepsilon}

\def \vnabla{\vec{\nabla}}

\begin{document}
\title{Electron-phonon mediated heat flow in disordered graphene}
\author{Wei Chen}
\affiliation{ Department of Physics, McGill University, Montreal, Canada H3A 2T8 }
\author{Aashish A. Clerk} \affiliation{
Department of Physics, McGill University, Montreal, Canada H3A 2T8 }

\begin{abstract}
We calculate the heat flux and electron-phonon thermal conductance in a disordered graphene sheet, going beyond a Fermi's Golden rule approach to fully account
for the modification of the electron-phonon interaction by disorder.
Using the Keldysh technique combined with standard impurity averaging methods in the regime $k_F l \gg 1$ (where $k_F$ is the Fermi wavevector, $l$ the mean free path),
we consider both scalar potential (i.e.~deformation potential) and vector potential couplings
between electrons and phonons.  We also consider the effects of electronic screening at the Thomas-Fermi level.  We find that the temperature dependence of the heat flux and
thermal conductance is sensitive to the presence of disorder and screening, and reflects the underlying chiral nature of electrons in graphene and the corresponding modification of their
diffusive behaviour.  In the case of weak screening, disorder enhances the low-temperature heat flux over the clean system (changing the associated power law from $T^4$ to $T^3$), and the deformation
potential dominates.  For strong screening, both the deformation potential and vector potential couplings make comparable contributions, and the low-temperature heat flux
obeys a $T^5$ power law.
\end{abstract}

\pacs{65.80.Ck, 72.10.Di, 44.10.+i}

\maketitle

\section{Introduction}

The potential to exploit the exceptional thermal properties of graphene in applications has recently generated considerable
activity~\cite{Levitov2011,Fong2012, Betz2012};  possible applications include sensitive bolometry and calorimetry for detecting infrared and THz radiation.  Such detectors
would ultimately be based on the simple heating of electrons in a graphene sheet by the absorption
of incident photons.  Ideally, the detector electrons would be thermally decoupled from their surroundings, thus allowing any heating produced by the incident radiation to be long-lived.  In this respect, graphene provides a potential advantage:
its low electron density and relatively weak electron-phonon coupling implies that the expected low-temperature thermal decoupling between electrons and the lattice~\cite{Ziman, Giazotto} could occur over a much wider temperature range than in a conventional metal~\cite{MacDonald2009, DasSarma2009}.  Further, one can effectively suppress the thermal link between graphene electrons and electrons in the contacts by employing superconducting leads~\cite{Roukes2000}.

Given the above, it is crucial to develop a rigorous and quantitative understanding of the thermal link between electrons and phonons in graphene at low temperatures.  Several recent theoretical works have addressed this problem in the case of clean graphene (i.e.~no electronic disorder)~\cite{Kubakaddi2009,DasSarma2009,MacDonald2009,Viljas2010}.  At low temperatures, the heat flux (per volume) between electrons and longitudinal acoustic phonons takes the general form:
\begin{equation}\label{eq:P}
P(T_{\rm{e}}, T_{\rm{ph}})=F(T_{\rm{e}})-F(T_{\rm{ph}})=\Sigma (T_{\rm{e}}^{\delta}-T_{\rm{ph}}^{\delta}),
\end{equation}
where $F(T)$ is called the energy control function,
 $\Sigma$ is a coupling constant, and $T_{\rm{e}}$ and $T_{\rm{ph}}$ are the temperatures of the electrons and lattice
(i.e.~phonons) respectively.  Previous works\cite{Kubakaddi2009,DasSarma2009,MacDonald2009,Viljas2010} find that $\delta = 4$ in the low-temperature limit, assuming an unscreened deformation potential
electron-phonon coupling.  This is identical to what would be expected for a clean conventional 2D metal~\cite{Viljas2010}.

In this work, we now ask how the above result is modified in the presence of electronic disorder.  While great experimental progress has been made in reducing disorder effects in
graphene \cite{DasSarmaReview}, the devices studied for bolometric applications in Refs.~\onlinecite{Fong2012, Betz2012} are sitting on a silicon substrate and have mean free paths that are $100$ nm or less.
In conventional metals, electronic disorder can strongly affect the electron-phonon coupling at temperatures low enough that the wavelength of a thermal phonon is comparable to (or longer than) the electronic mean free path~\cite{Altshuler, Reizer, Sergeev2000, Sergeev2005}.  This defines a characteristic temperature scale $T_{\rm dis}$ below which disorder effects are important,
\begin{equation}
	\kB T_{\rm dis} \equiv h s / l,
	\label{eq:Tdis}
\end{equation}
where $s$ is the speed of sound and $l$ is the electronic mean free path.
As discussed in Refs.~\onlinecite{Altshuler, Reizer, Sergeev2000, Sergeev2005} (and below), the effects of disorder are subtle:  depending on the nature of the disorder
and the electronic system, the power law $\delta$ in Eq.~(\ref{eq:P}) can either be enhanced by disorder or be suppressed~\cite{Altshuler, Reizer, Sergeev2000, Sergeev2005}.

Here, we study how the additional complexity arising from the unique electronic properties of graphene modify the interplay of disorder and the electron-phonon interaction.  The principle new ingredients arise from the effective chiral nature of carriers in graphene, which both modifies electronic diffusion, and allows for a new kind of effective vector-potential electron-phonon
coupling~\cite{Guinea1992, Ando2002,CastroNeto2009}. 
For simplicity, we will focus on disorder originating from charges in the substrate below the graphene flake, and thus take the disorder potential to preserve the symmetries (valley and sublattice) of the low-energy graphene Hamiltonian \cite{Falko2007, Falko2008}.  The impurity potential is thus also taken to be static, i.e.~it does not move with the graphene sheet.  We also consider the case where the graphene flake has been doped sufficiently that $k_F l \gg 1$ ($k_F$ is the Fermi wavevector), meaning that the standard impurity-averaged perturbation theory is appropriate.~\cite{Altshuler1980, Hikami1980}  Combining this approach with the Keldysh technique then allows us to rigorously address the electron-phonon interaction in the presence of disorder, in a manner analogous to the classic works looking at this physics in a conventional disordered metal~\cite{Altshuler, Reizer, Sergeev2000, Sergeev2005}.  We stress that properly addressing disorder effects involves going beyond the sort of Golden Rule calculation used to address the clean case \cite{Viljas2010}.

Note that since we work in the regime $k_F l \gg 1$, the temperature scale $T_{\rm dis}$ below which disorder effects emerge will
necessarily be well below the Bloch Gr\"uneisen temperature
$\kB T_{\rm BG} =  2 \hbar s k_F$; we will thus explicitly focus on temperatures $T < T_{\rm BG}$.
For a typical doped graphene electron density $10^{12}/ \mathrm{cm}^2$, $T_{\rm BG} \sim 70$K.
In contrast, for a typical mean free path of $100 $nm and a graphene acoustic phonon velocity
of $\sim 2\times10^4\,$m/s, $T_{\rm dis} \sim 10$K.
We note that a recent study examined disorder effects on electron-phonon interactions {\it above} the BG temperature~\cite{Levitov2011}.  Somewhat surprisingly, the expression derived in that work for impurity-assisted electron-phonon cooling for $T \gg T_{\rm BG}$ is exactly half of
 our expression for $e$-phonon cooling based on the deformation potential at $T \ll T_{\rm dis} < T_{\rm BG}$ (c.f. Table~\ref{tab:table1} and Eq.~(\ref{eq:ECDP3})).

\begin{widetext}

 \begin{center}
\begin{table}[positionspecifier]
	\centering
				\begin{tabular*}{1.0\textwidth}{@{\extracolsep{\fill}}p{2cm}*{6}{c}}
		\hline
		\hline
		\multicolumn{1}{c}{}&\multicolumn{4}{c}{Deformation potential}&\multicolumn{2}{c}{Vector potential}\\
		\hline
		\multicolumn{1}{c}{}&\multicolumn{2}{c}{$T<T_{\rm dis}$}& \multicolumn{2}{c}{$T_{\rm dis}<T<T_{BG}$} & \multirow{2}
			{*}{$T<T_{\rm dis}$}& \multirow{2}{*}{$T_{\rm dis}<T<T_{BG}$} \\
		\multicolumn{1}{c}{}& weak screening & strong screening & weak screening &
			\multicolumn{1}{c}{strong screening}  & \multicolumn{1}{c}{} & \multicolumn{1}{c}{} \\
		\hline
		\\
 			\multicolumn{1}{c}{$ F(T)  \frac{v_F^3 \rho_M}{E_F}$} &
			\multicolumn{1}{c}{$ \frac{2 \zeta(3)}{\pi^2}
			\frac{g^2_1  k^3_B }{  \hbar^4  l s^2} T^3$} &
		\multicolumn{1}{c}{$\frac{24 \zeta(5)}{ \pi^2}
			\frac{  g^2_1 \kB^5}{  \hbar^6  s^4 q^2_{\rm TF}l} T^5 $} &
		\multicolumn{1}{c}{$\frac{\pi^2}{15}
				\frac{g^2_1 k^4_B}{ \hbar^5  s^3}T^4$} &
		\multicolumn{1}{c}{$\frac{8\pi^4}{63}
			\frac{g^2_1 k^6_B}{  \hbar^7  s^5 q^2_{\rm TF}}T^6 $} &
		\multicolumn{1}{c}{$\frac{30  \zeta(5)}{ \pi^2}
			\frac{g^2_2 l k^5_B}{  \hbar^6  s^4}T^5$} &
		\multicolumn{1}{c}{$\frac{\pi^2}{15}\frac{g^2_2  k^4_B}{ \hbar^5  s^3}T^4 $} \\
		\\
		\hline
	\end{tabular*}
\caption{Energy control function of deformation potential and vector potential in graphene below the Bloch Gr\"uneisen temperature $ T_{\rm BG} =  2 \hbar s k_F/\kB$. $g_1$ and $g_2$ are the deformation potential and vector potential coupling constant respectively
(c.f.~Eqs.~(\ref{eqs:EPbarevertex})). $E_F$ is the Fermi energy with respect to the Dirac point, $v_F$ the Fermi velocity, $l$ the electronic mean free path, and
$\rho_M$ is the mass density of graphene per area.  $q_{\rm TF}$ is the Thomas-Fermi wavevector (as defined in Ref.~\onlinecite{VonOppen2009}),  and $\zeta(n)$ is the zeta function.}
	\label{tab:table1}
\end{table}
\end{center}

\end{widetext}

Our main results for the energy control function $F(T$) (c.f.~Eq.~(\ref{eq:P})) for graphene at low-temperatures are summarized in Table \ref{tab:table1}.  We consider the contribution to the electron-phonon heat flux arising from both the standard deformation potential coupling (DP), as well as from the effective vector potential coupling (VP).  In the absence of disorder (or at temperatures well above $T_{\rm dis}$ but below $T_{\rm BG}$), and in the absence of electronic screening, one finds that both these mechanisms contribute independently and in a similar manner:  the respective heat fluxes are each described by Eq.~(\ref{eq:P}) with $\delta = 4$ (in agreement with Refs.~\onlinecite{Kubakaddi2009,Viljas2010}).

In contrast, for $T < T_{\rm dis}$, the two coupling mechanisms are affected oppositely by the electronic disorder.  We find that the heat flux associated with the VP coupling is {\it suppressed} by disorder:  still neglecting screening, it is described now by Eq.~(\ref{eq:P})
with an enhanced power-law of $\delta = 5$.  Heuristically, this is attributed to the disorder-broadening of the graphene energy levels.
The effect of disorder on the heat flux associated with the DP coupling for $T < T_{\rm dis}$ is the opposite from the above:  {\it it is enhanced}.  It is described by a {\it reduced} power-law $\delta = 3$ (again, no screening), and will thus dominate the VP at low temperatures.  On a heuristic level, this enhancement is due to the diffusive charge dynamics, which effectively increases the time an electron interacts with a given phonon (i.e.~this becomes the time to diffuse across a phonon wavelength, as oppose the time needed to ballistically traverse this distance).  The absence of any diffusive enhancement of the VP coupling is (as we will show) a direct consequence of the non-conservation of pseudospin.

We also consider how including screening changes the above results; the importance of screening the $e$-ph interaction has been the subject of several recent
studies~\cite{DasSarma2011, VonOppen2009, VonOppen2010}.  As discussed extensively in Refs.~\onlinecite{VonOppen2009,VonOppen2010}, the DP coupling is expected to be screened, whereas the VP coupling is expected to be unscreened, as it induces no net electronic charge (i.e.~the effective
vector potential generated by a phonon field has opposite sign in the two graphene valleys).  As a result (see Table 1), even without disorder (i.e.~$T_{\rm dis} < T < T_{\rm BG}$), the DP and VP heat fluxes are not equivalent in the limit of strong screening:  the VP power-law remains $\delta = 4$, where the DP power law is increased to $\delta = 6$.  Similarly, at low temperatures where disorder effects matter, the VP power-law is unchanged, but the DP power law becomes $T^5$.

Note that our results suggest that even though the bare VP coupling strength $g_2$ is believed to be about an order-of-magnitude smaller than the bare DP coupling $g_1$
\cite{Ando2002}, if screening is strong,
its contribution to the heat flux could be comparable to or even large than that from the DP coupling.  This is despite the relative enhancement of the DP coupling over the VP coupling by disorder.
Further, our results suggest that the electron-phonon heat flux could be a means for empirically determining if screening is important.  In particular, the only way to obtain a $T^3$ power-law is via an unscreened DP coupling in the diffusive limit.  We note that measurements of the phonon contribution to the resitivity in a clean graphene sheet (as recently measured\cite{Kim2010}) cannot directly resolve this issue, as both DP and VP couplings contribute a $T^4$ dependence \cite{VonOppen2010}.

This paper is organized as follows. In Sec.~\ref{sec:Model}, we present our model and an outline of the calculational method.  This
includes a brief derivation of the kinetic equation of electrons in graphene in the Keldysh formalism (Sec.~\ref{subsec:Keldysh} ), as well as
a derivation and discussion of the diffusion propagator in graphene and the resulting diffusive renormalization of the electron-phonon vertex (Sec.~\ref{subsec:diffuson}).
We present the main results in Sec.~\ref{sec:RandD}, i.e. the heat flux due to electron phonon interaction in both weak and strong screening case. Finally, we briefly summarize the paper in
Sec.~\ref{sec:summary}.


\section{Model and calculation}\label{sec:Model}

\subsection{Model Hamiltonians of impurity and electron-phonon scattering}\label{subsec:Hamiltonian}

\subsubsection{Electrons in disordered graphene}

The low-energy electronic degrees of freedom are described by a massless Dirac Hamiltonian.
Focusing on a single valley
(the $\mathbf{K}_+$ valley), one has ~\cite{Wallace, Weiss, CastroNeto2009, DasSarmaReview}
\begin{equation}
	\HH	=
		\int d^2 \br \,
			\bpsi^{\dagger}(\br)
				\left(-i v_F \hat{\sigma}_j \partial_j +U(r) \hat{1} \right)
			\bpsi(\br)
	\label{eq:HDirac}
\end{equation}
where $v_F = 10^{6}m/s$ is the Fermi velocity, $\bpsi(\br) = \left(\begin{array}{c}
                                                          \psi_A(\br) \\
                                                           \psi_B(\br)
                                                        \end{array}\right)$
  is a spinor field operator describing the amplitude of electrons on the two sublattices,
$\hat{\sigma}_j$ ($j=x,y$) are Pauli matrices, and $U(\br)$ is  the disorder potential; we also set $\hbar = 1$ throughout unless otherwise indicated.  We do not include an index for spin or valley, as for the physics we consider, each spin and valley contributes in an equal and independent fashion.

As mentioned, we focus on a smooth disorder potential originating with impurities in the substrate.  We thus treat the impurity potential $U(\bf r)$ as a scalar potential with respect to both the valley degree of freedom and the sublattice degree of freedom (i.e.~pseudospin) \cite{Falko2007,Falko2008}.  In the standard way, $U$ will be treated as delta-correlated Gaussian disorder, with zero-mean and correlator:
\begin{equation}
	\langle U(\textbf{r})U(\textbf{r}')\rangle=w \delta(\textbf{r}-\textbf{r}').
\end{equation}
The corresponding scattering rate is $1/\tau \equiv v_F / l =\pi \nu w$, where $\nu=k_F/2\pi v_F$ is the density of states at the Fermi energy per spin per valley.

\subsubsection{Electron-phonon interaction}

The electron-phonon interaction in graphene has been studied extensively in Ref.~\onlinecite{VonOppen2009, VonOppen2010}. For suspended graphene, there are both in-plane phonon modes and flexural phonon modes (out-of-plane). In this work, we consider graphene on a substrate (as in recent experiments probing thermal properties~\cite{Fong2012, Betz2012}), such that flexural motion is suppressed; we thus only focus on in-plane motion.  Further, at low to moderate temperatures, the optical modes are barely excited and the dominant modes participating
in cooling of hot electrons are acoustic modes; we thus focus exclusively on the coupling to these modes.

Due to the Dirac Hamiltonian of electrons in graphene, there are two distinct electron-phonon coupling mechanisms~\cite{Guinea1992, Ando2002,CastroNeto2009}.
The first is a standard deformation potential coupling, which corresponds to a local dilation of the lattice.  In the Dirac theory, it appears
as a scalar potential (with respect to pseudospin).  The second mechanism is an effective gauge-field coupling or vector potential coupling.  This corresponds to the change in hopping matrix elements accompanying a pure shear deformation, and enters the Dirac theory the same way as an external gauge field (the only proviso being that this phonon-induced vector potential is valley-odd, and hence does not break time-reversal symmetry).

Letting $\bpsi_\bk$ denote a momentum-space electronic field operator and $b_{\eta,\bq}$ a phonon annihilation operator,
the total interaction between electrons and acoustic phonons can be written in the general form \cite{VonOppen2010}
\begin{equation}
	\HH_{ep}=\sum_{\eta = l,t}\sum_{\bk,\bq} \bpsi^\dag_{\bk+\bq} \, \hat{M}^\eta(\bq) \, \bpsi_{\bk} \left(b_{\eta,\bq}+b^\dag_{\eta,-\bq} \right).
	\label{eq:Hep}
\end{equation}
Here $\eta=l,t$ denote longitudinal (LA) and transverse (TA) acoustic modes respectively.  The $2 \times 2$ coupling matrices
$\hat{M}^\eta(\textbf{q})$ take the form:
\bse
	\label{eqs:EPbarevertex}
\begin{eqnarray}
	\hat{M}^l( \textbf{q} ) & = &
		i q \xi^l_{q}\left(\begin{array}{cc}
                                                      g_1 & -ig_2e^{2i\phi_{\bq}}\\
                                                      ig_2 e^{-2i\phi_{\bq}} & g_1
                                                    \end{array}\right),
                 	\label{eq:Lbare}	\\
	\hat{M}^t(\textbf{q}) & = &
			i q \xi^t_{q}\left(\begin{array}{cc}
                                                      0 & g_2e^{2i\phi_{\bq}}\\
                                                      g_2 e^{-2i\phi_{\bq}} & 0
                                                    \end{array}\right),
               \label{eq:Tbare}
\end{eqnarray}
\ese
where
\begin{equation}
	\xi^\eta_q=(\hbar/2 \rho_M\omega^\eta_q)^{1/2}.
\end{equation}
Here $\rho_M$ is the mass density of the graphene sheet, $\omega^\eta_q$ is the phonon frequency for $\eta$ mode, $\phi_{\bq}$ is the angle of the phonon wavevector $\textbf{q}$ with respect to the $x$ axis
(which is taken to be along the armchair direction of the graphene lattice). For simplicity, we take the speed of sound to be the same for LA and TA phonon and drop the superscript $\eta$ in the frequency from now on.  $g_1$ ($g_2$) is the deformation potential (vector potential) coupling constant.  Previous works have estimated
$g_1\sim 20-30$ eV and $g_2\sim 1.5$ eV\cite{Ando2002}, though we note that even the value of the deformation potential coupling is subject to some
debate~\cite{DasSarma2008,DasSarma2011,Jacobsen2012}.  Our theory is thus not tied to specific values of these parameters, and we keep both the DP and VP couplings in our discussion.
Note that transverse phonons induce only a pure shear deformation, and hence couple only through the vector potential.

\subsection{Keldysh formalism of the kinetic equation of electrons in graphene}\label{subsec:Keldysh}

Having established the basic electronic and electron-phonon Hamiltonians (c.f.~Eq.~(\ref{eq:HDirac}) and (\ref{eq:Hep})),
we now turn to our main goal of calculating the heat flux between electrons and phonons.  We consider the standard situation where each subsytem is independently in thermal equilibrium at its own temperature (electrons at $T_{\rm e}$, phonons at $T_{\rm ph}$).  In the disorder-free case, this heat flux can be conveniently calculated by using Fermi's Golden rule to calculate electron-phonon scattering rates~\cite{Kubakaddi2009, DasSarma2009, MacDonald2009,Viljas2010}.  Including disorder, we need a more general formalism, one that is capable of capturing the interference between electron phonon and electron impurity scatterings (i.e.~the vertex correction of the electron-phonon vertices $\hat{M}^\eta(\bq)$ by disorder). To that end, we make use of the Keldysh technique \cite{Kamenev2009,RammerBook}, coupled with standard disorder-averaged perturbation theory. Such an approach was used by Kechedzhi \emph{et al.} to study conductance fluctuations (in the absence of any electron-phonon coupling) \cite{Falko2008}.

We start by noting that the heat flux of interest (i.e. energy lost/gained by the electrons) can be directly related to the collision integral
$I_0(\ve,\textbf{p})$ appearing in a standard Boltzman equation describing the dynamics of the electronic phase-space distribution function $n[\ve,\bp; t]$:
\begin{equation}
	I_0(\ve,\textbf{p})  \equiv   \left[ \frac{d n[\ve,\bp]}{d t } \right]_{e-{\rm ph} \, \,  {\rm scatt}}.
\end{equation}
The collision integral $I_0(\ve,\textbf{p})$ tells us the rate of change of $n[\ve,\bp;t]$ due to the emission and absorption of acoustic phonons.

As electronic momentum relaxation is much faster than energy relaxation, to describe the latter process we can focus
on times longer than the electron momentum relaxation time; in addition, the distribution function will be sharply peaked on-shell (magnitude of momentum set by energy).  The relevant kinetics can thus be described
by an electron
distribution function $n(\ve;t)$  that depends only on energy, not on momentum.  The kinetic equation for the electronic distribution function $n(\ve;t)$ takes the form
\begin{equation}
	\frac{\partial n(\ve;t)}{\partial t}  =
	\bar{I}_0( \ve)  =
		\frac{1}{\pi\nu}  \int  \frac{d \mathbf{p} }{(2\pi)^2}
		I_0( \ve, \mathbf{p}) A(\ve, \mathbf{p}),
\end{equation}
where
\begin{equation}\label{eq:Spectrum}
	A(  \ve,\mathbf{p})= -\textrm{Im}\frac{2(\ve+\frac{i}{2\tau})}{(\ve+\frac{i}{2\tau})^2 -v^2_F p^2}
\end{equation}
is the electron spectral function.

The heat flux between electrons and lattice (for one valley and spin projection) is given by
\begin{equation}\label{eq:PDefn}
P(T_{\rm{e}},T_{\rm{ph}})=\nu\int d\ve \, \ve \bar{I}_0(\ve ).
\end{equation}

The collision integral $I_0(\ve,\textbf{p})$ is obtained in the standard manner by calculating the electronic Keldysh self-energies $\hSigma$ arising from the electron-phonon interaction, to first order.  One finds the general relation \cite{Falko2008} (see Appendix \ref{sec:KeldyshFormalism}):
\begin{eqnarray}
	\label{eq:Kinetic}
	&& I_0( \ve , \bp )  = \nonumber\\
	&&-\frac{i}{4} \textrm{Tr } \left[
		\hSigma^K +(1-2  n(\ve, \bp ; t))(\hSigma^A-\hSigma^R) \right].
\end{eqnarray}
The self-energies $\hSigma^j$ ($j = K, R$ and $A$) are $2 \times 2$ matrices (in pseudospin space), and are functions
of both energy $\ve$ and momentum $\bp$; they include the effects of disorder averaging.  As we are considering a quasi-equilibrium situation where both phonons and electrons are individually in thermal equilibrium, $n(\ve, \bp ; t)$
in Eq.~(\ref{eq:Kinetic}) can be replaced by a Fermi distribution function at temperature $T_{\rm e}$, and the self-energies can be calculated assuming phonons are in thermal equilibrium at temperature $T_{\rm ph}$.

\begin{figure}[ptb]
\includegraphics[width=7.0cm]{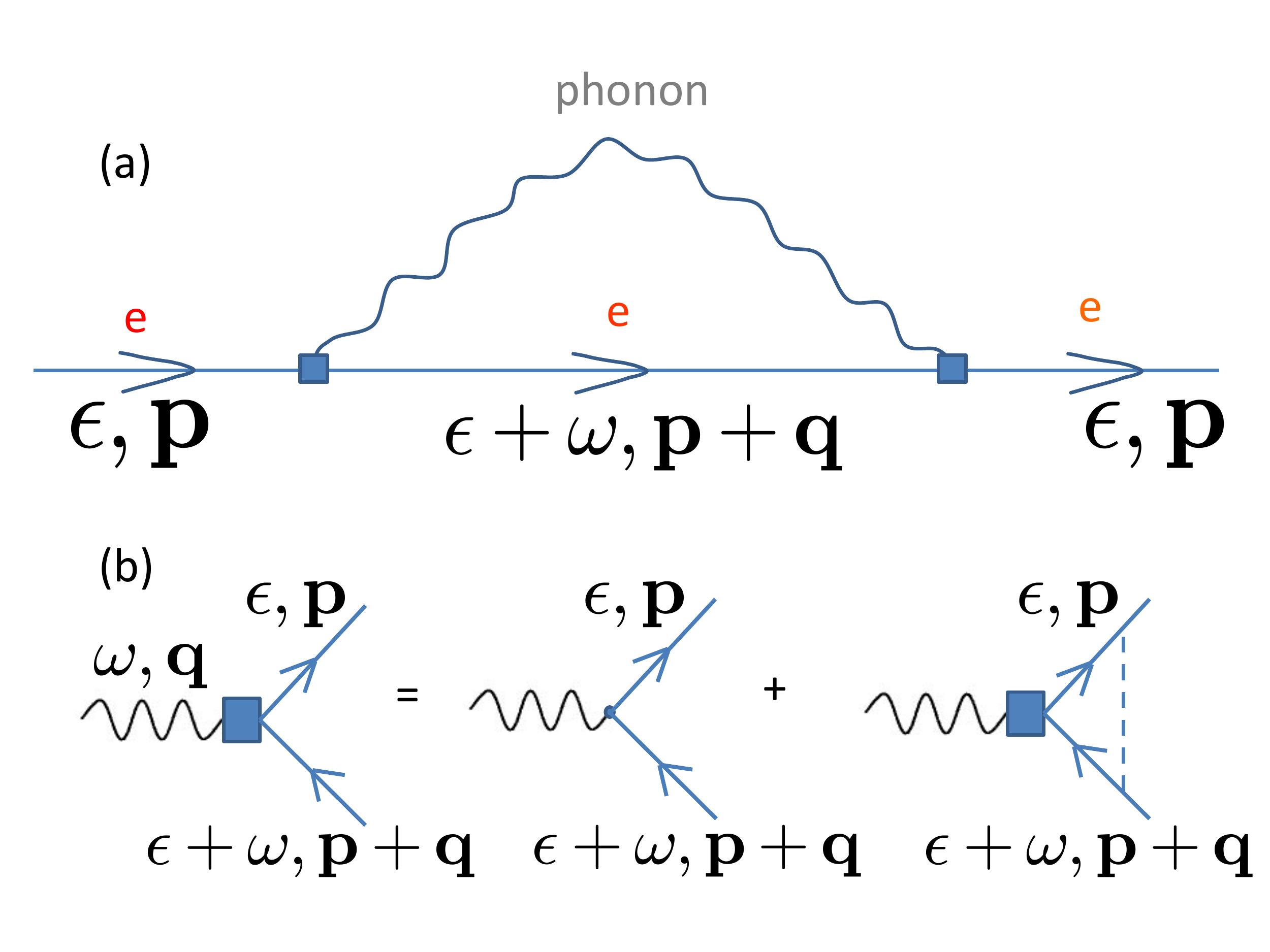}
\caption{Color online. (a)Electron self-energy diagram. The wavy line represents phonon propagator, the solid line represents electron propagator and the square block represents dressed electron phonon vertex by diffuson. (b) Vertex correction of the electron-phonon interaction by impurity scattering.  Dashed line represents an impurity average. The square block represents dressed vertex and the dot represents bare vertex.}\label{fig:self}
\end{figure}

In the regime of interest ($k_F l \gg 1$), the dominant self-energy diagram describing the leading-order electron-phonon contribution to the kinetic equation (in the presence of disorder) is shown in Fig.~\ref{fig:self}.  In this diagram,
the wavy line represents a phonon propagator $\Pi^\beta(\omega, \bq; \eta)$,
where $\beta = R, A, K$ denotes retarded, advanced and Keldysh propagators. The retarded and advanced phonon propagators appearing here are
\begin{equation}
\Pi^{R/A}(\omega, \textbf{q})=\frac{2\omega_q}{\omega^2-\omega^2_q\pm i\delta},
\end{equation}
with $\omega_q = s q$, while the Keldysh propagator is
\begin{equation}
	\Pi^K(\omega,\textbf{q}) = (1 + 2N(\omega,T_{\rm ph}) )
		\left(\Pi^R(\omega,\textbf{q}) -\Pi^A(\omega,\textbf{q}) \right),
\end{equation}
where $N(\omega,T_{\rm ph})$ is the Bose-Einstein distribution evaluated at $T = T_{\rm ph}$.

The solid line in Fig.~\ref{fig:self} represents an impurity-averaged electronic Green function $\hat{G}^\beta(\ve, \bp)$; note that these are $2 \times 2$
matrices in pseudospin space.  The retarded and advanced components are given by:
\begin{equation}
	\hat{G}^{R/A}(\ve, \textbf{p})=\frac{\ve\pm \frac{i}{2\tau}+v_F \vec{\sigma}\cdot \textbf{p}}{(\ve \pm \frac{i}{2\tau})^2-v_F^2\textbf{p}^2},
	\label{eq:GR}
\end{equation}
while the Keldysh electron Green function is
\begin{equation}
	\hat{G}^K(\ve, \textbf{q}) = \left(1- 2 n(\ve,T_{\rm e}) \right)
		\left( \hat{G}^R(\ve,\textbf{q})  -  \hat{G}^A(\ve,\textbf{q}) \right),
\end{equation}
where $n(\ve,T)$ is now the Fermi-Dirac distribution function evaluated at
$T = T_{\rm e}$ and chemical potential $E_F$, where $E_F$ is the Fermi energy (measured from the Dirac point).

Note that the electron Green function in Eq.~(\ref{eq:GR}) has an extremely simple form: it is just a free propagator with the substitution $\ve \rightarrow \ve  \pm i / 2 \tau$, corresponding to disorder-induced broadening of energy levels.  This broadening represents the first mechanism by which the
electron-phonon heat flux will be modified due to disorder; this broadening generally causes a {\it suppression}
of the heat flux.
The second key effect of disorder is via the vertex correction of the electron-phonon vertex appearing in the self-energy
 in Fig.~\ref{fig:self}.
Each vertex describes the emission or absorption of a phonon; heuristically, the disorder-induced vertex correction
corresponds to the modification of the amplitude of such a process due to the diffusive motion of electrons.
The full details of the renormalization of the $e$-phonon vertices by disorder in the Keldysh formalism in normal metals are presented in Ref.~[\onlinecite{Reizer}]. The Keldysh structure and renormalization of the $e$-phonon vertices in graphene can be treated in a similar fashion.  The only key difference comes from the
$2 \times 2$ matrix structure associated with pseudospin; as we will see, this leads to interesting new physical consequences.

The electron phonon vertices $\hat{M}$ (the $\eta$ superscript is dropped here) in the Keldysh technique are represented by the form $\hat{M}^{\gamma}_{\mu,\nu}$, where the upper index is for phonons and the lower electrons. The index $\gamma, \mu,\nu$ each has two components, $cl$ and $q$, due to the two-component structure of the electron and phonon fields in the Keldysh formalism\cite{Kamenev2009}. The vertices with different indices are renormalized differently by disorder in the Keldysh technique as shown in Ref.~\onlinecite{Reizer}. Here, we only present the simplest case, the renormalization of the vertex $\hat{M}^q_{cl,cl}$.  Upon summation of all self-energy diagrams, one finds that the form of this renormalized vertex appears directly
in the final expression for the collision integral, Eq.~(\ref{eq:collisionintegral}).  The renormalization of vertex $\hat{M}^q_{cl,cl}$ by disorder is depicted in Fig.~\ref{fig:self}b. We focus on this specific vertex in the remainder of this subsection and drop the upper and lower index from now on.

In general, we may describe the renormalization of a electron-phonon interaction vertex by
\begin{equation}
	\label{eq:RenormalizedEP}
	\hat{M}_{\rm diff}(\bq,\omega) =
		\check{D}(\bq,\omega) \circ \hat{M}_0(\bq)
\end{equation}
where $\hat{M}_0$ ($\hat{M}_{\rm diff}$) is the bare (renormalized) vertex, and
$\check{D}(\bq,\omega)$ is a linear operator acting in the space of $2 \times 2$ matrices. It represents the diffusion propagator for electrons in graphene,
with the non-trivial matrix structure reflecting the fact that charge and pseudospin diffusion are linked together.
It is convenient to write this expression using a basis of Pauli matrices. Defining the four-component vectors $\vec{m}, \vec{m}_{\rm diff}$ via
\bse\label{eqs:fourvector}
\begin{eqnarray}
	\hat{M}_0 & = & \vec{m}_0 \cdot \left( \hat{1}, \hat{\sigma}_x, \hat{\sigma}_y, \hat{\sigma}_z  \right),	\\
	\hat{M}_{\rm diff} & = & \vec{m}_{\rm diff} \cdot \left( \hat{1}, \hat{\sigma}_x, \hat{\sigma}_y, \hat{\sigma}_z  \right).
\end{eqnarray}
\ese
Eq.~(\ref{eq:RenormalizedEP}) takes the form:
\begin{equation}
	\label{eqs:PauliRep}
	\vec{m}_{\rm diff}(\bq,\omega) = \mathcal{D}(\bq,\omega) \cdot \vec{m}_0(\bq)
\end{equation}
where $\mathcal{D}$ is a $4 \times 4$ matrix.  Using this representation, the summation of ladder diagrams
depicted in Fig.~\ref{fig:self} results in the form:
\begin{eqnarray}\label{eq:DP}
	&&
		\mathcal{D}(\bq,\omega) =
			\left(1-\mathcal{P}(\bq,\omega) \right)^{-1},
		\nonumber\\
	&&
		\left[ \mathcal{P}(\bq,\omega) \right]_{\alpha\beta}=
		\nonumber\\
	&&\ \ \
		\frac{1}{2 \pi \nu \tau} \hat{\sigma}^\alpha_{ij}
			\int \frac{d^2\textbf{k}}{(2\pi)^2}
				\left[ \hat{G}^R(\bk+\bq,\omega) \right]_{ki}
				\left[ \hat{G}^A(\bk, 0)\right]_{jl}
			\hat{\sigma}^\beta_{lk}.\nonumber\\
\end{eqnarray}
Here, the indices $\alpha, \beta$ run from $0$ to $3$, and repeated indices are to be summed over;  we also use $\hat{\sigma}^0$ denote the $2 \times 2$ unit matrix and $\hat{\sigma}^i \ (i=1,2,3)$ are the Pauli matrices.
$\mathcal{P}$ describes a single ``rung" in a standard diffuson ladder.  Its $4 \times 4$ matrix structure is now directly
related to the fact that each propagator carries an initial and final charge or pseudospin index.  A similar structure is
encountered when considering diffusive dynamics in a system with strong spin-orbit coupling or in 2D helical metals, as studied by Burkov et al.~\cite{Burkov2004, Burkov2010}.

We will discuss the properties and physics encoded in the matrix diffusion propagator $\mathcal{D}$ in more detail in the next
subsection.  For now, we only show how it enters in the final expression for the collision integral (and hence the heat flux).
One finds that due to the causality structure of Keldysh Green functions, the two vertices in Fig.~\ref{fig:self}a cannot both be simultaneously dressed by impurity scattering.  Summing up all the self energy diagrams in the Keldysh formalism, one finally obtains:

\begin{widetext}
\begin{eqnarray}
	\label{eq:collisionintegral}
	 I(\ve, \textbf{p}) =
		-\frac{1}{2}\sum_{\eta = l,t} {\rm Tr   } \Bigg[  \int \frac{d\textbf{q}d\omega}{(2\pi)^3} R(\ve, \omega)
	  \left(  \left[  \hat{M}^\eta_0(\bq) \right]^\dag
		\Pi^R(-\omega; \eta)  \hat{G}^R(\ve+\omega, \textbf{p}+\textbf{q})
		\hat{M}^\eta_{\rm diff} (-\omega,-\textbf{q})  +  {\rm h.c.} \right)
		  \Bigg],
		\nonumber\\
\end{eqnarray}
\end{widetext}
where $R(\ve,\omega)$ is the expected combination of Bose-Einstein and Fermi-Dirac functions appropriate for
phonon emission and absorption processes:
\begin{eqnarray}
	R(\ve,\omega)  & = &
		N(\omega, T_{\rm ph}) n(\ve, T_{\rm e})(1-n(\ve + \omega, T_{\rm e}))
		\nonumber \\
	&&
		-(1+N(\omega, T_{\rm ph}))(1-n(\ve, T_{\rm e}))n(\ve+\omega, T_{\rm e})
		\nonumber \\
   &=& (n(\ve, T_{\rm e}) - n(\ve +\omega, T_{\rm e}))[N(\omega, T_{\rm{ph}})-N(\omega, T_{\rm{e}})]. \nonumber\\
\end{eqnarray}
The first term of $R(\ve,\omega)$ describes absorption of a phonon $\omega$ from energy state $\ve$ to $\ve+\omega$ and the second term describes emission of a phonon $\omega$ from energy state $\ve+\omega$ to state $\ve$.  As expected, $R(\ve, \omega)$ vanishes if $T_{\rm e} = T_{\rm ph}$.  We note that apart from the
matrix structure of the electron-phonon vertices and electron propagators, the expression for the collision integral has the same form
as that found for conventional diffusive metals\cite{Altshuler, Reizer, Sergeev2005}.
Nonetheless, we will see that the added matrix structure (which encodes the chiral nature of the graphene electronic excitations) gives rise to qualitatively new effects.

\subsection{Diffusion propagator and renormalization to the $e$-phonon vertex}\label{subsec:diffuson}

\subsubsection{Diffusion propagator}

It follows from Eqs.~(\ref{eq:PDefn}) and (\ref{eq:collisionintegral}) that a key part of the disorder-induced modification of the electron-phonon heat flux
is due to the modification of the effective electron-phonon interaction vertex.  This modification is in turn directly related to the chiral diffusive
dynamics of electrons in graphene, as described by Eq.~(\ref{eq:RenormalizedEP}).
In this subsection, we discuss the diffusion propagator in more details, as well
as the forms of the dressed electron-phonon vertices.  Note that we restrict our discussion here (as we do throughout the paper) on electrons
in the $\mathbf{K}_+$ valley.  While the sign of the chirality will be different for holes, or for the $\mathbf{K}_-$ valley, this sign
has no impact on the quantity of interest, the $e$-phonon heat flux in the presence of disorder.

We focus here on the most interesting diffusive regime, where  $ql\ll1, \omega\tau\ll 1$
(i.e., we are interested in length scales longer than $l$ and time scales longer than $\tau$). In this limit we can work
to lowest non-vanishing
order in $q l$ and $\omega \tau$.  The inverse diffusion propagator $\mathcal{D}^{-1}(\bq,\omega)
=1-\mathcal{P}(\bq,\omega)$ simplifies to
\begin{eqnarray}
	&\frac{1}{\tau}\mathcal{D}^{-1}(\bq, \omega)= \ \ \ \ \ \ \ \ \ \ \ \ \ \ \ \ \  \ \  \ \  \ \ \ \  \ \  \ \ \  \ \ \ \ \ \ \ \ \ \  \ \  \ \ \  \ \ \ \ \ \ \ \ \  \ \nonumber\\
	&\left(\begin{array}{cccc}
                               -i\omega +Dq^2 &0 & 0&0 \\
                              0& \frac{1}{2}(\frac{1}{\tau}-i\omega +Dq^2) & 0 & 0\\
                              0 & 0 & \frac{1}{2}(\frac{1}{\tau}-i\omega +Dq^2) & 0 \\
                              0& 0 & 0 & \frac{1}{\tau}
                            \end{array}\right) \nonumber\\
	&+\frac{1}{4}  \left(\begin{array}{cccc}
                                  0 & 2 iv_Fq_x & 2 iv_Fq_y  & 0\\
                                 2 iv_Fq_x & D(q^2_x-q^2_y) & 2 Dq_xq_y & 0 \\
                                  2 iv_Fq_y  & 2 Dq_xq_y & D(q^2_y-q^2_x)& 0 \\
                                  0& 0 & 0 & 0
                                \end{array}\right),\nonumber\\
        \label{eq:InvDMatrix}
\end{eqnarray}
in the diffusive limit, where $D=v_F l/2$ is the usual diffusion constant in two dimensions.

To gain intuition, it is useful to follow Ref.~\onlinecite{Burkov2004} and consider the real-space representation of the matrix diffusion propagator, which describes the coarse-grained
evolution of charge and pseudospin densities ($N(\br,t)$ and $S_j(\br,t)$ respectively, $j=x,y,z$).
The first term in Eq.~(\ref{eq:InvDMatrix}) would simply lead to uncoupled equations for each of these quantities:  $N$ would be described by a standard diffusion equation, while $S_x$ and $S_y$ would have an additional decay term (rate $1/\tau$), corresponding to the fact that pseudospin is not a conserved quantity.  $S_z$ has no dynamics in the limit we consider, as it precesses with frequency $E_F$ and averages away on the timescale $1 / \tau$.

The second term in Eq.~(\ref{eq:InvDMatrix}) complicates the above picture,
as it now links the dynamics of charge and pseudospin densities.  We thus obtain a set of
coupled diffusion equations, describing the the dynamics of these quantities (note that we have taken into account the fact that the Pauli matrices are twice of the pseudo-spin matrices):
\bse
\label{eqs:RealSpaceD}
\begin{eqnarray}
	&&
		\frac{\partial N}{\partial t} =
			D\nabla^2 N - v_F(\frac{\partial S_x}{\partial x}+\frac{\partial S_y}{\partial y})
		\label{eq:DiffEqN}, \\
	&&  \frac{\partial S_x}{\partial t} =
		\frac{3D}{2} \frac{\partial^2}{\partial x^2}S_x
		+\frac{D}{2}\frac{\partial^2}{\partial y^2}S_x
		+D\frac{\partial^2}{\partial x\partial y}S_y-\frac{S_x}{\tau}\nonumber\\
	&&
		-\frac{v_F}{2} \frac{\partial N}{\partial x},
	\label{eq:DiffEqSx} \\
	&&  \frac{\partial S_y}{\partial t}  =
		 \frac{D}{2}\frac{\partial^2}{\partial x^2}S_x
		+  \frac{3D}{2} \frac{\partial^2}{\partial y^2}S_x
		+  D\frac{\partial^2}{\partial x\partial y}S_x
		-\frac{S_y}{\tau}\nonumber\\
	&&
		 -\frac{ v_F}{2} \frac{\partial N}{\partial y}.
		 \label{eq:DiffEqSy}
\end{eqnarray}
\ese

These equations are analogous (but not identical) to the diffusive dynamics for charge and spin in a 2D helical metal \cite{Burkov2010} or the diffusion equations of Cooperons in graphene \cite{McCann2006}.  The interpretation here is similar to Ref.~\onlinecite{Burkov2010}:
the coupling between charge and pseudospin dynamics in the diffusive limit is a result of the effective helicity of the electronic eigenstates.
By helicity, we mean that at $\ve > 0$ eigenstate of Eq.~(\ref{eq:HDirac}), pseudospin will be aligned with momentum.  Thus, a positive gradient in say $S_x$ in the $x$-direction implies a corresponding positive gradient in density of electronic $x$-momentum.  This will then naturally cause the charge density $N$ to decrease in time:  this is the third term in Eq.~(\ref{eq:DiffEqN}).
Alternatively, writing Eq.~(\ref{eq:DiffEqN}) in the form of a continuity equation,
\begin{equation}
\frac{\partial N}{\partial t}=- \vnabla \cdot \textbf{J},
\end{equation}
one sees that the charge current density $\textbf{J}$ has the form
\begin{equation}
\textbf{J}=-D\vnabla N +v_F(S_x \hat{x}+S_y \hat{y}).
\end{equation}
The first term is the usual diffusive current, while the second term corresponds to a ``drift" current driven by the pseudospin density following from the Hamiltonian in Eq.~(\ref{eq:HDirac}) (i.e.~the current operator is the pseudospin operator).

Turning to the dynamics of pseudospin densities, Eqs.(\ref{eq:DiffEqSx}) and (\ref{eq:DiffEqSy}) again reflect the fact that pseudospin is not conserved, and effectively decays on a timescale $\tau$ due to elastic impurity scattering.  In addition, we see that the diffusion of these densities is
anisotropic:  this is also a simple consequence of helicity,  as a net pseudospin density in a specific direction also implies a net momentum density in this direction which reinforces the diffusion in this direction.

Finally, inverting Eq.~(\ref{eq:InvDMatrix}) (using as always the Pauli matrix representation defined in Eq.~(\ref{eqs:PauliRep})),
one finds the diffusion propagator in Eq.(\ref{eq:DP}) in the diffusive limit to be
\begin{eqnarray}\label{eq:DMatrix}
	&\mathcal{D}(\bq, \omega) =\ \ \ \ \ \ \ \ \ \ \  \ \ \ \ \ \ \ \ \ \ \  \ \ \ \ \ \ \ \ \ \ \ \ \ \ \ \ \  \ \ \ \ \  \ \ \ \ \nonumber\\
	 & \left(\begin{array}{cccc}
                        \frac{1}{(-i\omega+2Dq^2)\tau} & \frac{iq_xl}{(i\omega-2Dq^2)\tau} & \frac{iq_yl}{(i\omega-2Dq^2)\tau} & 0 \\
                        \frac{iq_xl}{(i\omega-2Dq^2)\tau} & \frac{3}{2}+\frac{i\omega\tau+(q^2_x-q^2_y)l^2}{2(i\omega-2Dq^2)\tau} &\frac{q_xq_yl^2}{(i\omega-2Dq^2)\tau} & 0 \\
                       \frac{iq_yl}{(i\omega-2Dq^2)\tau} & \frac{q_xq_yl^2}{(i\omega-2Dq^2)\tau} & \frac{3}{2}+\frac{i\omega\tau-(q^2_x-q^2_y)l^2}{2(i\omega-2Dq^2)\tau} & 0 \\
                        0 & 0 & 0 & 1
                      \end{array}\right).\nonumber\\
\end{eqnarray}
Note that the effective diffusion constant (i.e.~the coefficient of $q^2$ in the diffusion poles appearing above)
is {\it twice} the value of the standardly-defined $D$ appearing in Eqs.~(\ref{eqs:RealSpaceD}):
$D_{\rm{eff}} = 2 D  = v_Fl$.
This effective doubling of the diffusion constant is a direct consequence of the chiral nature of electrons in graphene, and is consistent with the results of previous transport studies \cite{Falko2007, Ando}.  Also note that as expected from our discussion following Eq.~(\ref{eq:InvDMatrix}), only the charge-charge component of $\mathcal{D}$ (i.e. the $(1,1)$ matrix element) diverges in the limit of small $\omega$ and $q$.  The lack of any corresponding large enhancement of the spin components of $\mathcal{D}$ is directly tied to the fact that pseudospin is not a conserved quantity.

\subsubsection{Disorder vertex correction of deformation-potential $e$-phonon vertex}

Having discussed the basic form of the matrix diffusion propagator, we now turn to the renormalization of the $e$-phonon interaction vertex due to diffuson,
as given by Eq.~(\ref{eqs:PauliRep}) and Eq.~(\ref{eq:DMatrix}).  Consider first
the DP contribution to the vertex.  The bare DP vertex is just a diagonal matrix in the sublattice basis (c.f.~Eqs.~(\ref{eqs:EPbarevertex})) proportional to the coupling constant $g_1$.  The impurity-dressed version takes the form:
\begin{equation}
	\label{eq:DPdiffusivelimit}
	\hat{M}_{\rm DP,diff }(\bq, \omega) =
			\frac{  i q \xi_q  g_1 }{\tau}
				\left(\begin{array}{cc}
                                               \frac{1}{(-i\omega+2Dq^2)} & -\frac{iqle^{-i\phi_{\bq}}}{(-i\omega+2Dq^2)} \\
                                               -\frac{iqle^{i\phi_{\bq}}}{(-i\omega+2Dq^2)} & \frac{1}{(-i\omega+2Dq^2)}
                                             \end{array}\right).
\end{equation}
The diagonal parts of the vertex acquire a diffusion pole protected by charge conservation, analogous to the case of a normal metal.
We will be interested in Eq.~(\ref{eq:DPdiffusivelimit}) with $\omega,q$ corresponding to a thermal phonon, $\hbar \omega = \hbar s q \simeq \kB T$.  As
$s \ll v_F$ in graphene, we thus have that over a wide range of temperature
\begin{equation}
	\label{eq:TRange}
	\frac{s}{v_F} T_{\rm dis} < T < T_{\rm dis},
\end{equation}
one has $\omega \ll D q^2$, and thus
the diagonal parts of the vertex in Eq.~(\ref{eq:DPdiffusivelimit}) will be enhanced by a factor $\sim 1 / (q^2 l^2$) compared to the clean case.  This corresponds to the well known diffusive enhancement of the electron-phonon interaction:  the diffusive motion effectively enhances the interaction time between an electron and a long-wavelength phonon.

More surprisingly, Eq.~(\ref{eq:DPdiffusivelimit}) implies that the diffusive renormalization of the deformation potential
induces a vector potential which is along the direction of wave vector $\textbf{q}$. This is a direct consequence of the chirality of the graphene electrons, which links
the dynamics of charge and pseudospin, and thus allows a scalar potential to generate a vector potential (i.e.~a potential which couples to pseudospin).  In the next section, we will show that this induced vector potential only gives a small contribution to the heat flux compared to the renormalized deformation potential coupling.

\begin{figure}[t]
\includegraphics[width=\columnwidth]{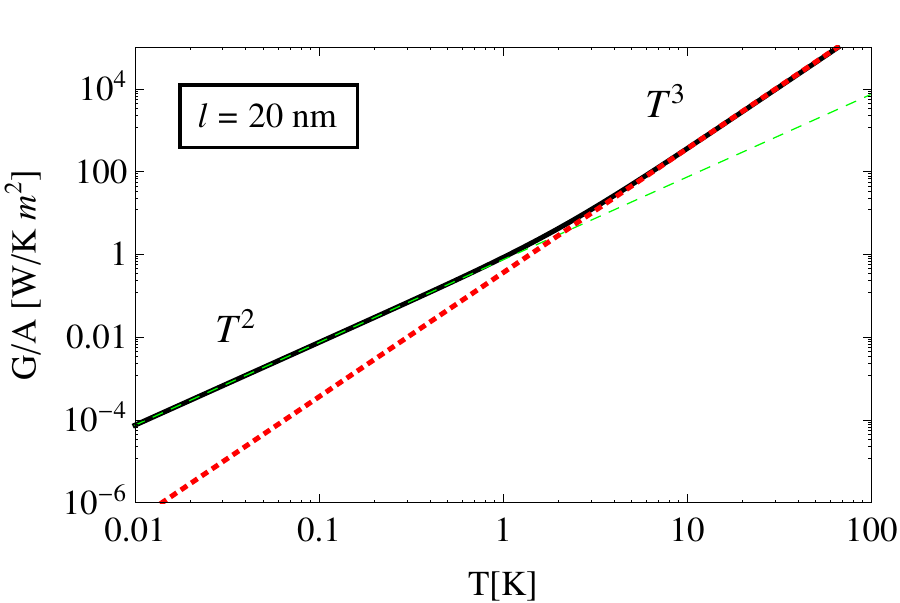}
\caption{Color online. Thermal conductance per unit area $G/A$ associated with the deformation potential coupling versus temperature $T$, including the effects of disorder, but without electronic screening.  We have
taken a bare coupling constant $g_1=20$~eV, carrier density $n=10^{12}/\rm{ cm}^2$ and mean free path $l=20 $nm. The black solid line is the full result of our theory.
The green-dashed line shows the asymptotic $T^2$ dependence in the low-temperature $T \ll T_{\rm dis}$ limit, whereas the red-dotted line shows the
asymptotic $T^3$ behaviour in the high-temperature (clean) limit.}  \label{fig:therconDP}
\end{figure}
\begin{figure}[t]
\includegraphics[width=\columnwidth]{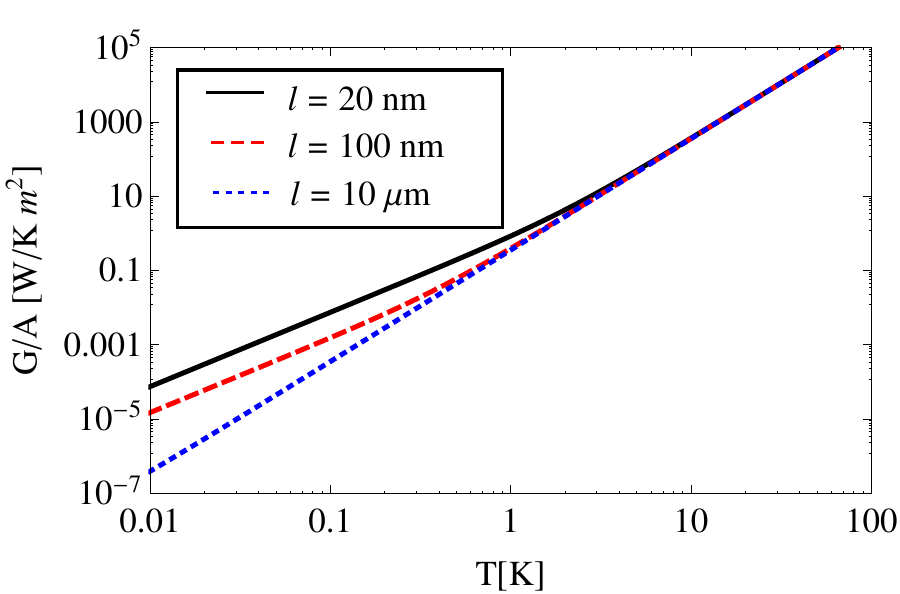}
\caption{Color online.
Thermal conductance per unit area $G/A$ associated with the deformation potential coupling versus temperature $T$,
including the effects of disorder, showing the effects of varying the mean-free path $l$ as
indicated; screening is neglected.  Remaining parameters are the same as Fig.~\ref{fig:therconDP}.  Both the enhancement of the low-temperature thermal conductance and shift of the
cross-over temperature with increasing disorder are clearly evident.}
\label{fig:therconDP2}
\end{figure}

\subsubsection{Disorder vertex correction of vector-potential $e$-phonon vertex}

The bare VP $e$-phonon vertex is given in Eqs.~(\ref{eqs:EPbarevertex}); it is purely off-diagonal in pseudospin space, and implies that a
phonon of wavector
$\bq = q \hat{\bq} = q (\cos\phi_\bq, \sin\phi_\bq)$ and polarization $\eta$
generates an effective vector potential $\mathbf{A}^l_{0}=g_2(\sin 2 \phi_\bq, \cos 2 \phi_\bq)$ ($\eta = l$) or
$\mathbf{A}^t_{0}=g_2(\cos 2 \phi_\bq, \sin 2 \phi_\bq)$ ($\eta = t$).
It is useful to decompose these vectors into their longitudinal and transverse parts:
\begin{equation}\label{eq:bareLAVP}
	\textbf{A}^\eta_{0}=A^\eta_{\parallel,0} \hat{\mathbf{q}} + A^\eta_{\perp,0} \hat{\mathbf{q}}_{\perp},
\end{equation}
where $\hat{\mathbf{q}}_{\perp}$ is a unit vector perpendicular to $\bq$.  One finds
\bse
\begin{eqnarray}
	A^l_{\parallel,0}=g_2\sin 3 \phi_\bq, \, \,\, \,  A^l_{\perp,0}=g_2\cos 3 \phi_\bq, \\
	A^t_{\parallel,0}=g_2\cos \phi_\bq, \, \, \, \, A^t_{\perp,0}=-g_2\sin \phi_\bq.
\end{eqnarray}
\ese

Given the linearity of the vertex correction described by Eq.~(\ref{eq:RenormalizedEP}), we can separately analyze how disorder changes the interaction with the transverse and longitudinal
phonon-induced vector potentials.  Each of these will yield an $e$-phonon vertex which is a $2 \times 2$ matrix in pseudospin space.  The bare vertices are:
\bse
\begin{eqnarray}
	\hat{M}^\eta_{\parallel,0}	& = &	A^\eta_{\parallel,0}\left( \cos \phi_\bq \hat{\sigma}_x + \sin \phi_\bq \hat{\sigma}_y \right),	\\
	\hat{M}^\eta_{\perp,0}	& = &	 A^\eta_{\perp,0} \left( -\sin \phi_\bq \hat{\sigma}_x + \cos \phi_\bq \hat{\sigma}_y \right).	
\end{eqnarray}
\ese
The corresponding renormalized vertices are given by Eqs.~(\ref{eqs:PauliRep}) and (\ref{eq:DMatrix}), yielding:
 \bse
 \label{eqs:RenormVP}
  \begin{eqnarray}
	\hat{M}^\eta_{\parallel,{\rm diff}}	& = &
		 \frac{2(i\omega-Dq^2)}{(i\omega-2Dq^2)} \hat{M}^\eta_{\parallel,0}
		 + \frac{iql / \tau}{i\omega-2Dq^2}  A^\eta_{\parallel,0} \hat{\sigma_0},\ \ \ \ \ \
	  	\label{eq:RNLVP}	\\
	\hat{M}^\eta_{\perp,{\rm diff}}	& = &
			2 \hat{M}^\eta_{\perp,0 }.
	\label{eq:RNTVP}
\end{eqnarray}
\ese
We see that the diffusive renormalization simply doubles the $e$-ph vertex associated with the transverse part of the phonon-induced vector potential; there is no diffusion pole here, as there is no charge associated with a transverse vector potential.  In contrast, the longitudinal part acquires a diffusion pole.  The vector potential part of the renormalized vertex is simply
the bare vertex multiplied by a factor of $\frac{2(i\omega-Dq^2)\tau}{(i\omega-2Dq^2)\tau}$.  This factor tends to 2 in the dc limit $ q\rightarrow 0, \omega\rightarrow 0$ and 1 in the regime we are most interested in in this work, i.e. $ql>s/v_F$. The renormalization factor of 2 in the dc limit is consistent with the renormalization of current vertex in graphene in the dc limit~\cite{Ando,Falko2008}.  Finally, we see that the vertex associated with the longitudinal vector potential also acquires a scalar potential (second term in Eq.~(\ref{eq:RNLVP}):  this is an induced deformation potential coupling, again arising from the helicity of electrons in graphene.

We stress that in contrast to the renormalized DP (c.f.~Eq.~(\ref{eq:DPdiffusivelimit})), Eqs.(\ref{eqs:RenormVP}) explicitly show that
there is {\it no} large enhancement of the VP vertex in the $\omega=0$ $q \rightarrow 0$ limit of interest.
As discussed, the lack of a diffusive enhancement is a direct consequence of the non-conservation of pseudospin and the consequent lack of a protected diffusion pole.  The net result is that the diffusive vertex correction discussed here {\it does not} significantly enhance the heat flux associated with the VP coupling at low temperatures.

Finally, for completeness, we give the full form of the dressed VP $e$-phonon vertex.  For the interaction with LA phonons,
combining
Eqs.~(\ref{eq:bareLAVP}) to (\ref{eqs:RenormVP}) yields:
\begin{widetext}
\begin{eqnarray}\label{eq:VPdiffusivelimit}
	 \hat{M}^{\eta=l}_{{\rm VP},\rm diff}(\bq, \omega)  =  i q \xi^l_q g_2
		\left(\begin{array}{cc}
                        -\frac{i q l}{-i \omega \tau + q^2 l^2} \sin{3\phi_{\bq}}  &
                     			-i \left(  \frac{3}{2} + \frac{i \omega \tau / 2}{ i \omega \tau - q^2 l^2  }  \right)   e^{ 2 i \phi_\bq }
						- i \frac{ q^2 l^2 / 2 }{ -i \omega \tau + q^2 l^2  } e^{ -4 i \phi_\bq } \\
		i \left(  \frac{3}{2} + \frac{i \omega \tau / 2}{i \omega \tau - q^2 l^2  }  \right)   e^{ -2 i \phi_\bq }
						+ i \frac{ q^2 l^2 / 2 }{ -i \omega \tau + q^2 l^2  } e^{ 4 i \phi_\bq }  &
                       			 -\frac{i q l}{-i \omega \tau + q^2 l^2} \sin{3\phi_{\bq}}
                \end{array}\right).
    \nonumber\\
\end{eqnarray}
\end{widetext}
The full interaction vertex for TA phonons can be obtained in a similar fashion.  One finds that vector potentials arising from TA and LA phonons make identical contributions to the heat flux.

\section{Results and Discussions}\label{sec:RandD}

\subsection{Heat flux without screening}

Having now determined both the renormalized electron-phonon vertices (c.f.~Eqs.~(\ref{eq:DPdiffusivelimit}),
(\ref{eq:VPdiffusivelimit})) as well as the disorder-averaged
electronic Green functions (c.f.~Eq.~(\ref{eq:GR})), we have all the necessary ingredients to evaluate Eq.~(\ref{eq:collisionintegral})
for the electronic collision integral.  From this, Eq.~(\ref{eq:PDefn})
directly yields the desired electron-phonon heat flux.  As with the disorder-free case, we again find that the DP and VP couplings contribute independently; we can thus meaningfully discuss the flux associated with each coupling.  It is useful to express each of these heat fluxes in terms of an energy control function $F_\alpha(T)$ ($\alpha = {\rm DP}, {\rm VP}$), defined via:
\begin{equation}
	\label{eq:FDefn}
	P_{\alpha}(T_{\rm{e}}, T_{\rm{ph}})
		=			\nu\int d\ve \, \ve
				\bar{I}_{{\alpha},0}(\ve)
					\equiv F_{\alpha}(T_{\rm{e}})-F_{\alpha}(T_{\rm{ph}}).
\end{equation}
Here, $\bar{I}_{\alpha,0}(\ve)$ is the momentum-average of the collision integral corresponding to the coupling mechanism $\alpha$.

We discuss each mechanism in turn, focusing as always on the regime $\frac{s}{v_F  } T_{\rm dis} < T \ll T_{BG}$; as discussed, the lower limit here allows us to ignore the frequency dependence of the renormalized electron-phonon vertices.  We also first discuss our results in the absence of any electronic screening.

\subsubsection{Deformation potential heat flux}

The energy control function determining the DP heat flux through the whole temperature regime $\frac{s}{v_F  } T_{\rm dis} < T \ll T_{BG}$ is obtained from Eq.~(\ref{eq:CollisionIntegralDP}) and Eq.~(\ref{eq:HFDP3}) to be
\begin{widetext}
\begin{equation}
	F_{\rm DP}(T)  =
		 4 g_1^2 \frac{s}{v_F} \frac{\nu  }{2 \pi \rho_M}  \int_0^\infty dq q^3  \,
		\left[
			q l \left( \frac{1}{\sqrt{1+q^2 l^2}}  +   \frac{1}{q^2 l^2} \right)
			  -  \frac{1}{q l}
				\left(1-\frac{1}{\sqrt{1+q^2 l^2}} \right)
		\right]  N(\omega_q,T).
		\label{eq:FDP}
\end{equation}
\end{widetext}
Here, $\omega_q = s q$, and we have included an overall factor of $4$ reflecting the identical contribution from both valleys and both physical spin projections.  The integrand reflects the contribution from phonons having momentum $q$ to the heat flux.
The first term corresponds to the contribution stemming only from the diagonal parts of the renormalized DP electron-phonon vertex, whereas the second term corresponds to off-diagonal terms (i.e.~the effective vector potential generated by the chiral diffusion).  The clean limit can easily be obtained by taking $l \rightarrow \infty$, yielding:
\begin{equation}
	F_{\rm DP, clean} \equiv \lim_{l \rightarrow \infty} F_{\rm DP}(T)= \frac{\pi^2}{15} g^2_1 \frac{E_F }{\hbar^5 \rho_M  v^3_F s^3} \left( \kB T \right)^4,
\end{equation}
where we explicitly include factors of $\hbar$ in the expression. This result is the same as the heat flux for deformation potential in clean graphene obtained in previous theoretical work.\cite{Viljas2010, MacDonald2009, Kubakaddi2009}

For finite $l$, both terms in Eq.~(\ref{eq:FDP}) contribute.  For long-wavelength phonons satisyfing $ql \ll 1$, the first term
dominates the second (effective vector potential) term by a large factor $1 / (ql)^2$.  In contrast, for phonons
having $ql \sim 1$, both terms make comparable contributions.
In the diffusive limit $T < T_{\rm dis}$, the energy control function reduces to
\begin{equation}\label{eq:ECDP3}
	F_{\rm DP}(T)	=	
		\frac{2 \zeta(3) }{\pi^2} g^2_1  \frac{ E_F }{ \hbar^4 \rho_M  s^2 v^3_F l} \left( \kB T \right)^3.
\end{equation}
The power law becomes $T^3$ instead of $T^4$; this is the result of the diffusive enhancement of the diagonal
parts of the renormalized DP vertex, in complete analogy to what happens in a conventional disordered metal~\cite{Altshuler, Reizer, Sergeev2000, Sergeev2005}.
The corresponding thermal conductance $G = (d / d \Delta) P(T+\Delta,T)$ associated with the DP coupling is shown in Figs.~\ref{fig:therconDP} and \ref{fig:therconDP2}.

As noted in the introduction, the expression in Eq.~(\ref{eq:ECDP3}) is exactly twice of that of the $e$-phonon
heat flux associated with impurity-assisted ``supercollisions", as described by by Song {\it et al.} in
Ref.~\onlinecite{Levitov2011}, for temperatures $T \gg T_{\rm BG}$.  Unlike our study, Ref.~\onlinecite{Levitov2011} considers $e$-phonon scattering dressed by only a {\it single} impurity scattering event, and simply sums the effect of each impurity.  In our approach, the $q$-dependence of the integrand in Eq.~(\ref{eq:FDP}) in the dirtly limit $l \rightarrow 0$ can ultimately be traced to the diffusive enhancement of the DP vertex (c.f.~Eq.~(\ref{eq:DPdiffusivelimit})).  In contrast, the corresponding $q$-dependence in Ref.~\onlinecite{Levitov2011} can be traced to the energy of a virtual electronic state in a second-order process involving both $e$-phonon and $e$-impurity scattering events.


\subsubsection{Vector potential coupling heat flux}

\begin{figure}[t]
\includegraphics[width=\columnwidth]{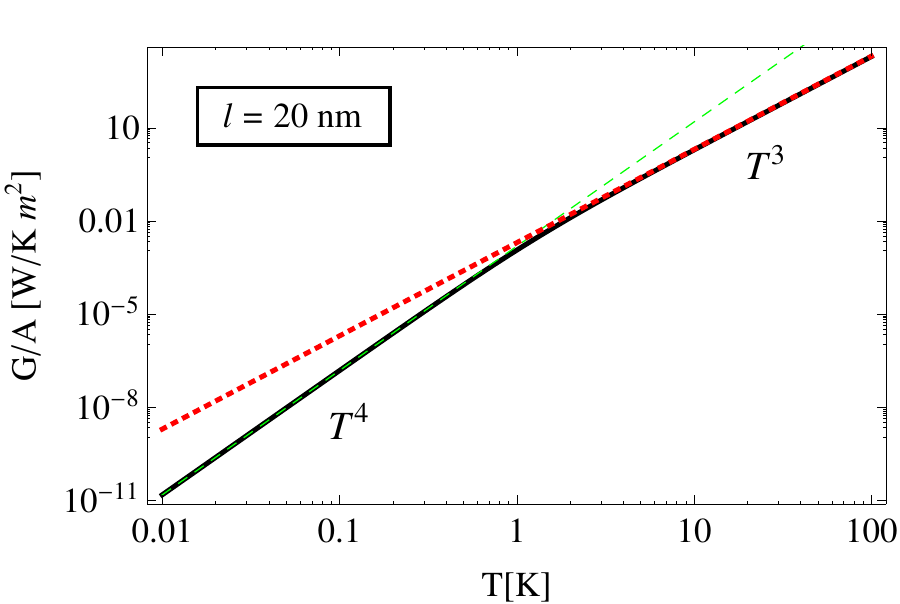}
\caption{Color online.
Thermal conductance per unit area $G/A$ associated with the vector potential coupling versus temperature $T$, including the effects of disorder, but without electronic screening.  We have
taken a bare coupling constant $g_2=1.5$~eV, carrier density $n=10^{12}/\rm{ cm}^2$ and mean free path $l=20 $nm. The black solid line is the full result of our theory.
The green-dashed line shows the asymptotic $T^4$ dependence in the low-temperature $T \ll T_{\rm dis}$ limit, whereas the red-dotted line shows the
asymptotic $T^3$ behaviour in the high-temperature (clean) limit.}  \label{fig:therconVP}
\end{figure}

\begin{figure}[ptb]
\includegraphics[width=\columnwidth]{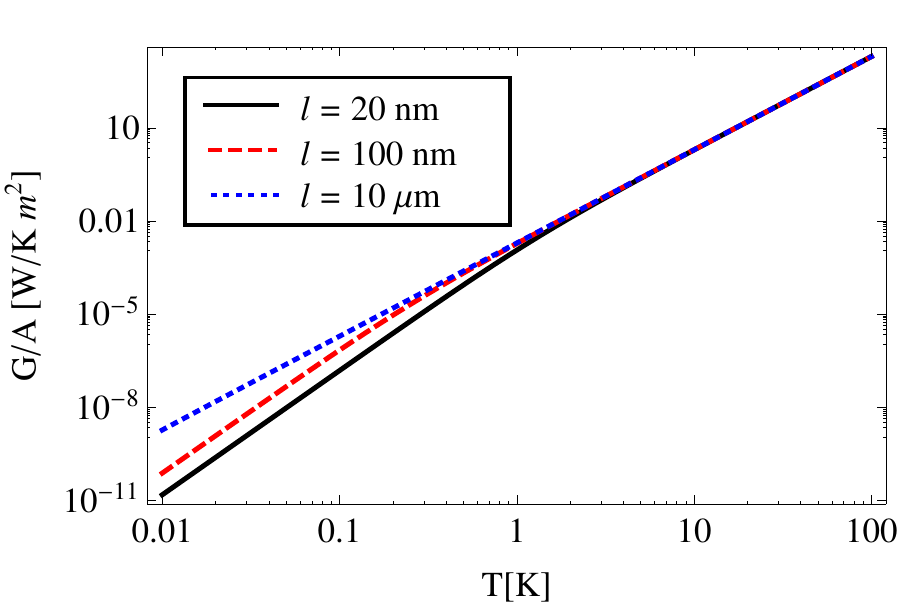}
\caption{Color online.
Thermal conductance per unit area $G/A$ associated with the vector potential coupling versus temperature $T$, including the effects of disorder, showing the effects of varying the mean-free path $l$ as
indicated; screening is neglected.  Remaining parameters are the same as Fig.~\ref{fig:therconVP}.  Both the suppression of the low-temperature thermal conductance and shift of the
cross-over temperature with increasing disorder are clearly evident.}
\label{fig:therconVP2}
\end{figure}

The energy control function $F_{\rm VP}(T)$ for vector potential coupling in the temperature regime $\frac{s}{v_F  } T_{\rm dis} < T \ll T_{BG}$ is obtained from Eq.~(\ref{eq:CollisionIntegralVP}) and Eq.~(\ref{eq:HFVP}) to be
\begin{widetext}
\begin{eqnarray}\label{eq:FVP}
F_{\rm VP}(T)=4 g_2^2 \frac{s}{v_F} \frac{\nu  }{2 \pi \rho_M}  \int dq  q^3
\left[ql\left(\frac{1}{\sqrt{1+q^2l^2}}+\frac{1}{2(1+q^2l^2)}\right)-\frac{1}{2q l}(1-\frac{1}{\sqrt{1+q^2 l^2}})\right]N(\omega_q,T).\nonumber\\
\end{eqnarray}
\end{widetext}
We have again included an overall factor of 4 in front to take into account the valley and spin degeneracy.
The first term in the square bracket is the contribution from the vector potential after renormalization and the second term is the contribution from the induced deformation potential coupling from vector potential.
At $ql\ll1$, the second term gives a contribution about $1/6$ of the first term, while at $ql\gg 1$, the second term is smaller by a factor of $1/2ql$. 
In the clean limit $T \gg T_{\rm dis}$, the energy control function for vector potential has the same functional form as that for the DP coupling, i.e.,
\begin{equation}\label{eq:HFVP4}
	F_{VP}(T)=\frac{\pi^2}{15} g^2_2 \frac{E_F }{\hbar^5 \rho_M v^3_F s^3} \left( \kB T \right)^4.
\end{equation}
In the opposite diffusive limit $T \ll T_{\rm dis}$, the energy control function for the VP coupling reduces to
\begin{equation}\label{eq:ECVP5}
	F_{VP}(T)	=
		\frac{30 \zeta(5)}{\pi^2} g^2_2  \frac {E_F l }{\hbar^6 \rho_M  s^4 v^3_F} \left( \kB T\right)^5.
\end{equation}
Unlike the DP coupling, we see disorder increases the power of temperature of the low-temperature heat flux, indicating a suppression of heat flux from VP coupling.  As discussed, the lack
of a diffusive enhancement of the VP coupling is directly tied to the non-conservation of pseudospin; the main remaining effect of disorder  is a simple broadening of the
electronic eigenstates.  The result is that for $T \ll T_{\rm dis}$, disorder suppresses VP-mediated heat transport.  Note that as the bare VP coupling constant $g_2$ has been estimated to be
more than an order-of-magnitude smaller than the corresponding DP coupling constant $g_1$\cite{Ando2002}, it follows that in absence of screening, the heat flux associated with the VP coupling is expected to be negligible in comparison to that associated with the DP coupling.
The thermal conductance associated with the VP coupling is shown in Figs.~\ref{fig:therconVP} and \ref{fig:therconVP2}.

\subsection{Heat flux with electronic screening}

We now consider how the above results are altered if one includes the screening of the $e$-phonon interaction.  As discussed extensively by von Oppen et al \cite{VonOppen2009}, the deformation potential coupling will be subject to screening at long wavelengths in the usual manner, whereas the vector potential will not be screened, as it does not induce any
net charge density. In this work, we are interested in temperatures such that $T >  (s / v_F) T_{\rm dis}$ (c.f.~Eq.~(\ref{eq:TRange})), implying that dynamic screening effects (which are also sensitive to disorder\cite{Sergeev2000}) will be unimportant.  Further, in the temperature regime of interest ($T < T_{\rm BG}$), a simple Thomas-Fermi approach to screening is expected to suffice \cite{DasSarma2011}.  The result is that the bare DP coupling constant $g_1$ in Eqs.~(\ref{eqs:EPbarevertex}) now becomes dependent on the
magnitude of the phonon wavevector $q$:
\begin{equation}
	g_{1,\rm{sc}}(q)= g_1 \frac{q}{q+q_{TF}},
\end{equation}
where the Thomas-Fermi wavector $q_{\rm TF}$ is given by \cite{DasSarmaReview,DasSarma2011}
\begin{equation}
	q_{\rm TF} = 4 \frac{e^2}{\kappa \hbar v_F}  k_F
\end{equation}
and $\kappa$ is an effective dielectric constant.  Using the value of $\kappa$ appropriate to graphene on a Si${\textrm O}_2$ substrate, one has $q_{\rm TF} \simeq 3.2 k_F$ \cite{DasSarmaReview}.
Note that as we focus on the regime $k_F l \gg 1$, the effects of screening will generally set in at a much higher temperature
$T_{\rm sc} = s q_{\rm TF} / \kB$ than the temperature $T_{\rm dis}$ below which
disorder-effects becomes important.  We note that a recent experiment measuring the $e$-phonon contribution to the electrical resistivity of a suspended graphene flake suggests that
screening does not seem to be playing a role even when $T < T_{\rm sc}$ \cite{Kim2010}, as the results are compatible with the predictions for an unscreened deformation potential interaction (see also Ref.~\onlinecite{DasSarma2011}).

One can now easily include the effects of screening into our theory by making the substitution $g_1 \rightarrow g_{1,{\rm sc}}(q)$ in Eq.~(\ref{eq:FDP}) for the energy control function $F_{\rm DP}(T)$ determining the deformation-potential mediated heat flux.  One finds:
\begin{equation}
	F_{\rm DP, sc}(T) = \begin{cases}
		\frac{ 8\pi^4g^2_1 E_F k^6_B}{63\rho_M \hbar^7 v^3_F s^5 q^2_{TF}}T^6,
			& \mbox{if } T_{\rm dis} \ll T \ll T_{\rm sc},  \\
		\frac{24 \ g^2_1 E_F}{\pi^2 \rho_M \hbar^6 v^3_F s^4 q^2_{TF}l}\zeta(5) T^5,
			& \mbox{if } T \ll T_{\rm dis} , T_{\rm sc}.
	\end{cases}
\end{equation}

\begin{figure}[t]
\includegraphics[width=\columnwidth]{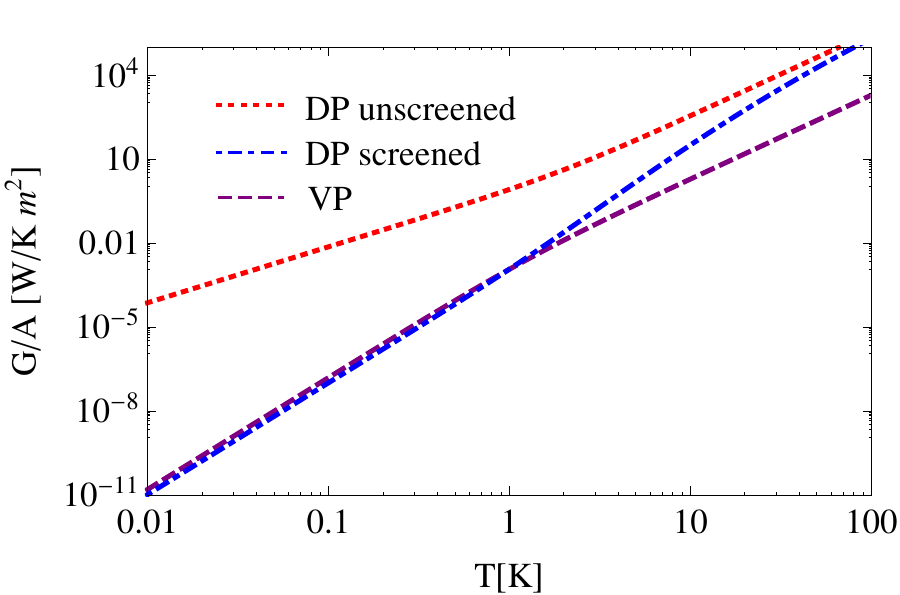}
\caption{Color online. Thermal conductance per unit area $G/A$ versus temperature $T$, showing the effects of electronic screening.   $g_1=20$~eV, $g_2 = 1.5$~eV, a carrier density $n=10^{12}/\rm{ cm}^2$ and a mean free path $l=20$ nm.  The red short-dashed line correspond to an unscreened deformation-potential coupling, while the blue dashed-dotted line corresponds to a screened deformation-potential coupling, with a Thomas-Fermi wavevector $q_{\rm TF} = 3.2 k_F$ as appropriate for graphene on Si$\textrm{O}_2$ \cite{DasSarmaReview}.  The dashed
purple curve is the contribution from the vector potential coupling.  Despite its much smaller bare coupling constant, we see that at low temperatures, both the deformation potential
and vector potential couplings make almost equal contributions when both screening and disorder effects are included.}
\label{fig:ScreeningDirty}
\end{figure}

\begin{figure}[t]
\includegraphics[width=\columnwidth]{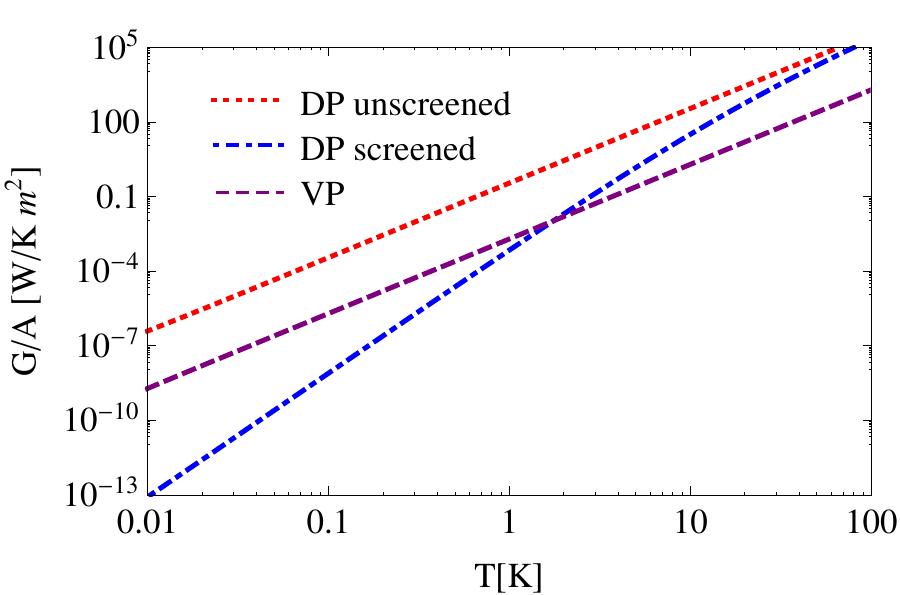}
\caption{Color online. Thermal conductance per unit area $G/A$ versus temperature $T$, showing the effects of electronic screening.
Parameters are identical to Fig.~\ref{fig:ScreeningDirty}, except we have now taken the mean free path to be $l = 10 \, \mu$m, meaning that we are effectively
in the clean limit.  Unlike the disordered case shown in Fig.~\ref{fig:ScreeningDirty}, we now see that at low temperatures, the contribution of the
vector potential coupling dominates that from the deformation potential coupling.}
\label{fig:ScreeningClean}
\end{figure}

The suppression of the DP coupling by screening at low temperatures $T \ll T_{\rm dis}$ implies that its associated heat flux can now become comparable or even smaller in magnitude to that associated with the VP coupling, c.f~Eq.~(\ref{eq:ECVP5}).  In the low temperature limit, both mechanisms yield energy control functions $F(T) \propto T^5$, with:
\begin{equation}
	\frac{ F_{\rm DP, sc}(T) }{ F_{\rm VP}(T)} \sim
		\frac{g^2_1}{g^2_2}\cdot\frac{1}{q^2_{TF}l^2}.
\end{equation}
We see that the largeness of $q_{\rm TF} l$ can compensate for the relative smallness of $g_2$ with respect to $g_1$, leading both mechanisms to make comparable contributions.  This behaviour is demonstrated in Fig.~\ref{fig:ScreeningDirty}, where the thermal conductance versus temperature for both mechanisms is presented, for both the strongly and weakly-screened cases.  The fact that both mechanisms are comparable is markedly different from what happens in the screened, disorder-free case, which is realized when $T_{\rm dis} \ll T \ll T_{\rm sc}$.  In this case,
the VP heat flux will dominate the DP heat flux at low temperatures, as it scales like $T^4$ as opposed to $T^5$ (see Fig.~\ref{fig:ScreeningClean}).  This behaviour in the clean limit is similar to expectations for $e$-phonon contribution to the electrical resistivity, where it has also been argued that the VP coupling can dominate at low temperatures \cite{VonOppen2010}.

Finally, we note that with our theory, the only way to obtain a $T^3$ power law in the heat flux at low temperatures (i.e.~$\delta = 3$ in Eq.~(\ref{eq:P})) is via an unscreened deformation potential.  Thus, measurements of the low temperature heat flux could also serve as a diagnostic tool for assessing the importance of screening the deformation potential.

\section{Conclusions}\label{sec:summary}

We have presented a comprehensive theory showing how electronic disorder modifies the electron-phonon interaction in graphene (both the vector potential and deformation potential couplings), and how this in turn has observable consequences for the heat flux between the electrons and lattice (acoustic) phonons.  We focused on the relatively simple situation where the graphene is doped away from the Dirac point, and where the impurity potential can be considered smooth on atomic scales (implying that the disorder potential preserves the pseudospin and valley symmetries of the graphene Hamiltonian).  We found that the unusual diffusion dynamics of electrons in graphene that results from their chirality also has implications for how disorder modifies electron-phonon physics.  We also found that this modification is quite different for the deformation potential coupling versus the effective vector potential coupling.  In the absence of screening, the
contribution to the heat flux in Eq.~(\ref{eq:P}) from both couplings has temperature dependence of $T^4$ in the clean limit $T\gg s/l$, consistent with previous work.
 In the disorder limit $T<s/l$, however, disorder affects the two types of couplings differently and the power law of heat flux for the two couplings becomes different. We found that the total effect of disorder enhances the heat flux from DP coupling, however, it suppresses the heat flux from VP coupling. The power law in Eq.~(\ref{eq:P}) for DP coupling becomes $T^3$ in the disorder limit while for VP coupling, the power law becomes $T^5$ in the same limit without screening.  Thus, without screening, the DP coupling is expected to dominate heat transport in both the clean limit and disorder limit, given that DP coupling is believed to be much stronger than vector potential coupling.

We also considered the effects of screening on the above physics, adopting the Thomas-Fermi approximation.  Only the DP is affected by screening; it suppresses it, and thus at low temperatures, its contribution to the heat flux scales like $T^6$ in the clean limit and $T^5$ in the disordered limit.  Without disorder, one would thus expect the VP to dominate at low temperatures due to the screening of DP, similar to expectations for the electron-phonon contribution to the resistivity in clean graphene sheets \cite{VonOppen2010}.  However, when disorder effects are included along with screeening, we find that both the DP and VP coupling mechanisms can make comparable contributions at low temperature.

In the future, it would be interesting (though challenging) to extend these results to situations of lower doping (where the Fermi energy is closer to the Diract point), and to include a richer structure of disorder that can break the symmetries of the clean graphene Hamiltonian (as has been done in, e.g.~, studies of weak-localization \cite{Falko2007}).

\acknowledgments
We acknowledge useful discussions with Anton V. Andreev, K.C. Fong, Tami Pereg-Barnea and K. C. Schwab.  This work was supported by NSERC and the Canadian Institute for Advanced Research (CIFAR).

\appendix

\section{Keldysh formalism of the kinetic equation}\label{sec:KeldyshFormalism}
The Green's function in the Keldysh space is a matrix as
\begin{equation}
\hat{G}=\left(\begin{array}{cc}
        \hat{G}^K & \hat{G}^R \\
        \hat{G}^A & 0
      \end{array}\right),
\end{equation}
where $\hat{G}^R$ and $\hat{G}^A$ are the retarded and advanced Green's function respectively. The Keldysh component $\hat{G}^K$ in general could be parameterized as $\hat{G}^K=\hat{G}^R\circ F-F\circ \hat{G}^A$, where $F$ is the Hermition distribution function matrix. The Green function obeys the following Dyson's equation:
\begin{equation}
\left(\begin{array}{cc}
  0 & (\hat{G}_0^A)^{-1}-\hat{\Sigma}^A \\
  (\hat{G}_0^R)^{-1}-\hat{\Sigma}^R & -\hat{\Sigma}^K
\end{array}\right)
\circ \left(\begin{array}{cc}
                                \hat{G}^K & \hat{G}^R \\
                               \hat{G}^A & 0
                              \end{array}\right)=1,
\end{equation}
where the bare retarded and advanced Green function in graphene are $(\hat{G}_0^{R/A})^{-1}(x,x')=\delta(x-x')(i\partial_{t'}+i\hat{\vec{\sigma}}\cdot\vec{\partial}_{r'})$ in the sublattice basis. $\hat{\Sigma}^{R/A}$ and $\hat{\Sigma}^K$ are the retarded, advanced and Keldysh component of the electron self energy respectively. The circle $\circ$ indicates integration over intermediate coordinates.

The equation for the Keldysh component reads
\begin{equation}\label{eq:Keldysh}
F\circ (\hat{G}_0^A)^{-1}-(\hat{G}_0^R)^{-1}\circ F=\hat{\Sigma}^K+F\circ \hat{\Sigma}^A-\hat{\Sigma}^R \circ F.
\end{equation}

The distribution function matrix $F(x, x')$ in graphene could be decomposed to four components,
\begin{equation}
F(x, x')=\sum^3_{i=0} f_i(x,x')\hat{\sigma}^i,
\end{equation}
where $\hat{\sigma}^0$ is the $2\times 2$ unit matrix and $\hat{\sigma}^i, i=1,2,3$ are the Pauli matrices.

The Wigner transformation of Eq.~(\ref{eq:Keldysh}) gives the kinetic equation of the distribution function as:
\begin{eqnarray}\label{eq:Keldysh_WT}
&&-\sum^3_{i=0} i\hat{\sigma}^i\partial_{\tau}f_i(t, \rho; \ve, \textbf{p}) -2iv_F\sum^3_{i,j,k=1}\epsilon_{ijk}\hat{\sigma}^k  k_j f_i(t, \rho; \ve, \textbf{p})  \nonumber\\
&&=\hat{\Sigma}^K(\ve, \textbf{p})+F(t, \rho; \ve,\textbf{p}) \hat{\Sigma}^A(\ve, \textbf{p})-\hat{\Sigma}^R(\ve, \textbf{p}) F(t, \rho; \ve, \textbf{p})\nonumber\\
&&=I(\ve, \textbf{p};t),
\end{eqnarray}
where the function $f_0(t, \rho; \ve, \textbf{p})=1-2n(t, \rho; \ve, \textbf{p})$ and $n(t, \rho; \ve, \textbf{p})$ is the charge density distribution function, $f_i(t, \rho; \ve, \textbf{p})$, $i=1,2,3$ represent the pseudo-spin density distribution and $\epsilon_{ijk}$ is the three-dimensional anti-symmetric tensor. The parameters $t$ and $\rho$ are center of mass time and coordinates respectively.
 Since the translation symmetry in space is restored after averaging over impurities, we drop the $t$ dependence in the text.

The right hand side of this equation is the collision integral $I(\ve,\textbf{p}; t)$ (times i) in presence of interactions. $\hat{\Sigma}^R, \hat{\Sigma}^A, \hat{\Sigma}^K$ are respectively the retarded, advanced and Keldysh components of the electron self energy due to $e$-phonon and impurity scatterings. From the left hand side of the above equation, one finds that the pseudo-spin density distribution functions $f_{i=1,2,3}$ are small in a factor of $1/E_F \tau$ compared to $f_0$. To leading order, one can replace $F(t, \rho; \ve,\textbf{p})$ on the right hand side by the scalar charge component $f_0(\ve,\textbf{p}; t)$.

 The collision integral is a $2\times 2$ matrix in the sublattice basis, which can be decomposed to components of $\hat{\sigma}^0$ and $\hat{\sigma}^i \ (i=1,2,3)$. The three components of $\hat{\sigma}^i \ (i=1,2,3)$ give a measure of the pseudo-spin density induced by interaction, while the $\hat{\sigma}^0$ component determines the collision integral for the charge density distribution function and is the one of interest in this paper. From Eq.(\ref{eq:Keldysh_WT}), one gets the collision integral for charge distribution function
\begin{eqnarray}
&I_0(\ve, \textbf{p})=\frac{\partial n(\ve,\textbf{p};t)}{\partial t}=-\frac{1}{2}\frac{\partial f_0(\ve, \textbf{p};t)}{\partial t}\ \ \ \ \ \ \ \ \ \ \ \ \ \ \ \ \ \ \ \ \ \ \ \nonumber\\
&=-\frac{i}{4} \textrm{Tr} \left[
		\hat{\Sigma}^K(\ve, \textbf{p})+(1-2  n(\ve,\textbf{p}; t))(\hat{\Sigma}^A-\hat{\Sigma}^R)(\ve, \textbf{p}) \right]\nonumber\\
\end{eqnarray}
as shown in Eq.~(\ref{eq:Kinetic}).

\section{Full form of the diffusion propagator and renormalized $e$-phonon vertex}\label{sec:DPandvertex}
The full form of the diffusion propagator $\mathcal{D}(\omega, \textbf{q})$ crossing the whole temperature regime $\frac{s}{v_F} T_{\rm dis} < T \ll T_{BG}$ is quite complicated, yet in the temperature regime $T>\frac{s}{v_F}T_{dis}$, the frequency dependence of the diffusion propagator can be dropped and the diffusion propagator in Eq.~(\ref{eq:DP}) is simplified to
\begin{eqnarray}
&&\mathcal{D}(\omega\rightarrow 0, \textbf{q})=\nonumber\\
&&\left(\begin{array}{cccc}
                                     1+\frac{\sqrt{1+q^2l^2}}{q^2l^2} & -\frac{i}{ql}\cos{\phi_{\bq}} & -\frac{i}{ql}\sin{\phi_{\bq}} & 0 \\
                                     -\frac{i}{ql}\cos{\phi_{\bq}} & 1+\frac{1-\cos{2\phi_{\bq}}}{2\sqrt{1+q^2l^2}} & -\frac{\sin{2\phi_{\bq}}}{2\sqrt{1+q^2l^2}} & 0\\
                                     -\frac{i}{ql}\sin{\phi_{\bq}} &  -\frac{\sin{2\phi_{\bq}}}{2\sqrt{1+q^2l^2}} & 1+\frac{1+\cos{2\phi_{\bq}}}{2\sqrt{1+q^2l^2}} & 0\\
                                     0 & 0 & 0 & 1
                                   \end{array}\right)\nonumber\\
\end{eqnarray}
in the Pauli matrix basis.

The four vector representation of the bare deformation potential vertex in the Pauli matrix basis is $\vec{m}_{\rm{DP},\ 0}=iq\xi^l_q(1,0,0,0)g_1$. The renormalized vertex according to Eq.~(\ref{eqs:PauliRep}) in such basis becomes
\begin{equation}
\vec{m}_{\rm{DP},\ \rm{diff}}(\omega\rightarrow 0, \textbf{q})=iq\xi^l_q\left(\begin{array}{c}
                  1+\frac{\sqrt{1+q^2l^2}}{q^2l^2} \\
                  -\frac{i}{ql}\cos{\phi_{\bq}}  \\
                 -\frac{i}{ql}\sin{\phi_{\bq}}\\
                 0
               \end{array}\right)^\mathcal{T} g_1,
\end{equation}
 where the superscript $\mathcal{T}$ means transpose of the column vector to row vector and the same for $\vec{m}_{\rm{VP},\ \rm{diff}}$ below. Written in the sublattice basis, the renormalized deformation potential vertex becomes
\begin{eqnarray}
&&\hat{M}_{\rm{DP}, \rm{diff}}(\omega\rightarrow 0, \textbf{q})\nonumber\\
&&=iq\xi^l_q\left(\begin{array}{cc}
                       1+\frac{\sqrt{1+q^2l^2}}{q^2l^2} & -\frac{ie^{-i\phi_{\bq}}}{ql} \\
                       -\frac{ie^{i\phi_{\bq}}}{ql} & 1+\frac{\sqrt{1+q^2l^2}}{q^2l^2}
                     \end{array}\right)g_1.
\end{eqnarray}
In the diffusive limit $\omega\tau\ll1, ql\ll1$, it reduces to Eq.~(\ref{eq:DPdiffusivelimit}) (dropping the frequency dependence there); while in the clean limit $ql\gg1$, it reduces to the bare vertex.

The four vector representation of the bare vector potential for LA phonon in Eq.~(\ref{eqs:EPbarevertex}) is
\begin{equation}
\vec{m}_{\rm{VP},0}=iq\xi^l_q\left(0,
                \sin{2\phi_{\bq}}, \cos{2\phi_{\bq}},
                0
              \right)g_2
\end{equation}
in the Pauli matrix basis. The renormalized vertex according to Eq.~(\ref{eqs:PauliRep}) then becomes
\begin{eqnarray}
&&\vec{m}_{\rm{VP},\rm{diff}}(\omega\rightarrow 0, \textbf{q})\nonumber\\
&&=iq\xi^l_q\left(\begin{array}{c}
                       -\frac{i\sin{3\phi_{\bq}}}{ql} \\
                       \sin{2\phi_{\bq}}(1+\frac{1}{2\sqrt{1+q^2l^2}})-\frac{\sin{4\phi_{\bq}}}{2\sqrt{1+q^2l^2}} \\
                       \cos{2\phi_{\bq}}(1+\frac{1}{2\sqrt{1+q^2l^2}})+\frac{\cos{4\phi_{\bq}}}{2\sqrt{1+q^2l^2}} \\
                       0
                     \end{array}\right)^\mathcal{T} g_2\nonumber\\
\end{eqnarray}
in the same basis. Written in the sublattice basis, it reads
\begin{widetext}
\begin{eqnarray}
&&\hat{M}_{\rm{VP},\rm{diff}}(\omega\rightarrow 0, \textbf{q})\nonumber\\
&&=iq\xi^l_q\left(\begin{array}{cc}
                                     -\frac{i\sin{3\phi_{\bq}}}{ql}& -ie^{2i\phi_{\bq}}(1+\frac{1}{2\sqrt{1+q^2l^2}})-i\frac{e^{-4i\phi_{\bq}}}{2\sqrt{1+q^2l^2}} \\
                                     ie^{-2i\phi_{\bq}}(1+\frac{1}{2\sqrt{1+q^2l^2}})+i\frac{e^{4i\phi_{\bq}}}{2\sqrt{1+q^2l^2}}  & -\frac{i\sin{3\phi_{\bq}}}{ql}
                                   \end{array}\right)g_2.
\end{eqnarray}
\end{widetext}
In the diffusive limit, it reduces to Eq.~(\ref{eq:VPdiffusivelimit}) (again dropping the frequency dependence there); while in the clean limit $ql\gg1$, it reduces to the bare vertex.

\section{Details of the matrix collision integral}
\label{sec:heatflux}
\subsubsection{Deformation potential}
The renormalized deformation potential in the whole regime of $ql$ is presented in Appendix B. Plugging in the renormalized deformation potential to Eq.~(\ref{eq:collisionintegral}) and integrating over the phonon frequency, one gets the collision integral for deformation potential coupling as
\begin{widetext}
\begin{eqnarray}\label{eq:CI}
I_{\rm{DP},0}(\ve, \textbf{p})&=&\frac{\partial n(\ve, \textbf{p};t)}{\partial t}=i\int \frac{d\textbf{q}}{(2\pi)^2}g^2_1\{(q\xi_q)^2\frac{1 }{(\ve +\omega_q +\frac{i}{2\tau})^2-v^2_F|\textbf{p}+\textbf{q}|^2}\nonumber\\
&&\left[[(\ve+\omega_q+\frac{i}{2\tau})(1+\frac{\sqrt{1+q^2l^2}}{q^2l^2})
+i\frac{v_F}{l}+
iv_F l\frac{ q_x p_x+q_y p_y}{q^2 l^2}]+h.c.\right]R(\ve,\omega_q)-(\omega_q\rightarrow -\omega_q, \textbf{q}\rightarrow -\textbf{q})\},\nonumber\\
\end{eqnarray}
\end{widetext}
where $\omega_q=sq$.

The collision integral for deformation potential after average over the electron momentum becomes
\begin{widetext}
\begin{eqnarray}\label{eq:CollisionIntegralDP}
&&\bar{I}_{\rm{DP},0}(\ve)=\frac{1}{\pi\nu}\int\frac{d\textbf{p}}{(2\pi)^2}A(\ve, \textbf{p})I_{\rm{DP},0}(\ve, \textbf{p})\nonumber\\
&&=\tau\int \frac{d\textbf{q}}{(2\pi)^2}g_1^2\{q^2\xi^2_q
\left[(1+\frac{\sqrt{1+q^2 l^2}}{q^2 l^2})\frac{1}{\sqrt{1+q^2 l^2}}-\frac{1}{q^2 l^2}(1-\frac{1}{\sqrt{1+q^2 l^2}})\right]R(\ve, \omega_q)-(\omega_q\rightarrow -\omega_q, \textbf{q}\rightarrow -\textbf{q})\}.\nonumber\\
\end{eqnarray}
\end{widetext}

The heat flux is then
\begin{eqnarray}\label{eq:HFDP3}
P_{DP}(T_{\rm{e}}, T_{\rm{ph}})=\nu\int d\ve \ve \bar{I}_{\rm{DP},0}(\ve)=F_{DP}(T_{\rm{ph}})-F_{DP}(T_{\rm{e}})\nonumber\\
\end{eqnarray}
where the energy control function $F_{DP}(T)$ is presented in Sec.~\ref{sec:RandD}.

\subsubsection{Vector potential}
Plugging in the renormalized vector potential in Appendix B to the collision integral Eq.~(\ref{eq:collisionintegral}) and separating the component for the charge distribution function, one gets the kinetic equation of the charge distribution function due to vector potential coupling after average over the angel of electron momentum as
\begin{widetext}
\begin{eqnarray}\label{eq:CollisionIntegralVP}
\frac{\partial n(\ve; t)}{\partial t}=\bar{I}_{\rm{VP},0}(\ve)&=&\tau\int \frac{d\textbf{q}}{(2\pi)^2} g^2_2\{q^2 \xi^2_q
[(1+\frac{1}{2\sqrt{1+q^2l^2}})\frac{1}{\sqrt{1+q^2l^2}}-\frac{1}{2q^2 l^2}(1-\frac{1}{\sqrt{1+q^2 l^2}})]R(\ve, \omega_q)\nonumber\\
&&-(\omega \rightarrow -\omega_q, \textbf{q} \rightarrow -\textbf{q})\}.
\end{eqnarray}
\end{widetext}

The heat flux due to vector potential coupling is
\begin{equation}\label{eq:HFVP}
P_{\rm{VP}}(T_{\rm{e}}, T_{\rm{ph}})=\nu\int d\ve\ve \bar{I}_{\rm{VP},0}(\ve)=F_{\rm{VP}}(T_{\rm{e}})-F_{\rm{VP}}(T_{\rm{ph}}),
\end{equation}
where the energy control function $F_{\rm{VP}}(T)$ is presented in Sec.~\ref{sec:RandD}.

\bibliography{bibtex_graphene_paper}

\end{document}